\documentclass[12pt]{iopart}
\usepackage{amssymb,graphicx,bm,xcolor}
\definecolor{r0}{rgb}{1,0,0}
\definecolor{g0}{rgb}{0,0.5,0}
\definecolor{y0}{rgb}{0.75,0.75,0}
\mathchardef\mhyphen="2D
\begin{document}

\title{Flat-top plasma operational space of the STEP power plant}

\author{E Tholerus$^1$, F J Casson$^1$, S P Marsden$^1$, T Wilson$^1$, D~Brunetti$^1$, P Fox$^1$, S J Freethy$^1$, T C Hender$^1$, S~S~Henderson$^1$, A Hudoba$^1$, K~K~Kirov$^1$, F Koechl$^2$, H~Meyer$^1$, S I Muldrew$^1$, C Olde$^1$, B~S~Patel$^1$, C M Roach$^1$, S~Saarelma$^1$, G Xia$^1$ and the STEP team$^1$}

\address{$^1$ CCFE, Culham Science Centre, Abingdon, Oxon, OX14 3DB, UK}
\address{$2$ ITER Organization, Route de Vinon-sur-Verdon, CS 90\,046, 13067 St. Paul Lez Durance Cedex, France}
\ead{emmi.tholerus@ukaea.uk}

\begin{abstract}
STEP is a spherical tokamak prototype power plant that is being designed to demonstrate net electric power. The design phase involves the exploitation of plasma models to optimise fusion performance subject to satisfying various physics and engineering constraints. A modelling workflow, including integrated core plasma modelling, MHD stability analysis, SOL and pedestal modelling, coil set and free boundary equilibrium solvers, and whole plant design, has been developed to specify the design parameters and to develop viable scenarios. The integrated core plasma model JETTO is used to develop individual flat-top operating points that satisfy imposed criteria for fusion power performance within operational constraints. Key plasma parameters such as normalised beta, Greenwald density fraction, auxiliary power and radiated power have been scanned to scope the operational space and to derive a collection of candidate non-inductive flat-top points. The assumed auxiliary heating and current drive is either from electron cyclotron systems only or a combination of electron cyclotron and electron Bernstein waves. At present stages of transport modelling, there is a large uncertainty in overall confinement for relevant parameter regimes. For each of the two auxiliary heating and current drive systems scenarios, two candidate flat-top points have been developed based on different confinement assumptions, totalling to four operating points. A lower confinement assumption generally suggests operating points in high-density, high auxiliary power regimes, whereas higher confinement would allow access to a broader parameter regime in density and power while maintaining target fusion power performance. 
\end{abstract}

%
% Uncomment for keywords
\vspace{2pc}
\noindent{\it Keywords}: STEP, integrated modelling, flat-top, JINTRAC, spherical tokamak
%
% Uncomment for Submitted to journal title message
\submitto{\NF}

\section{Introduction}\label{sec:intro}
\subsection{The spherical tokamak concept}
STEP (Spherical Tokamak for Energy Production) is a DEMO-class~\cite{demo} fusion ex\-pe\-ri\-ment, intended to demonstrate net electricity production using the spherical tokamak (ST) concept, with start of operation targeting 2040~\cite{step,step3,muldrew}. The main feature that distinguishes STs from conventional tokamaks is the small aspect ratio, which maximises the relative extent of regions with favourable curvature along closed magnetic field lines. This allows for operation at higher $\beta$ while avoiding pressure driven MHD instabilities such as ballooning modes. Thus, lower magnetic field strength can be used for the same kinetic pressure compared to conventional tokamaks. In addition, the lower aspect ratio also increases the natural elongation of the plasma~\cite{menard3}. As the bootstrap current fraction increases with both (poloidal) $\beta$ and elongation \cite{wilson,bootstrap}, less power is required for current drive in STs, improving the fusion $Q$ for a steady-state device. Conventional tokamaks require large sizes and magnetic field strengths to reach reactor relevant confinement times, largely because of the expected negative $\beta$ scaling of the confinement time when expressed in dimensionless ``physics'' parameters \cite{iter,costley}. However, experiments on spherical tokamaks such as NSTX and MAST have indicated a weak negative or even positive scaling of energy confinement with respect to $\beta$ \cite{buxton}, which would allow access to reactor relevant conditions at much smaller major radii, assuming that the suggested scaling laws extrapolate to these regimes.

The weaker magnetic field, lower power, and potentially smaller sizes required for reactor relevant conditions could make STs a generally more cost effective alternative to conventional tokamaks for commercial fusion power with steady state, non-inductive operation. However, the compact size leads to significant engineering challenges, particularly in the central column. The narrow central columns required by STs could limit the location of heating and fuelling systems mostly to the outer (low-field) side of the vessel. It also sets an upper limit to currents in the toroidal field coils and the central solenoid, which have to be sufficiently protected from radiation and mechanical stress. The more compact design relative to conventional tokamaks also puts higher requirements on the power load handling for the divertor and other plasma facing components. Exhaust is already one of the major technological challenges for conventional tokamak reactors.

\subsection{STEP design parameters}
STEP intends to demonstrate a net steady-state electric power output of at least 100\,MW, corresponding to an engineering fusion energy gain factor $Q_\mathrm{eng} = P_\mathrm{out}/P_\mathrm{in}$ just above 1.0 ($P_\mathrm{out}$ is the gross electrical output, whereas $P_\mathrm{in}$ is the electrical power required to drive core systems and auxiliary heating during steady-state operation). The main geometric parameters planned for the presently envisaged STEP design is a major radius $R_\mathrm{geo} = 3.6$\,m (radius at the geometric centre of the last closed flux surface in the midplane), an aspect ratio $A  = \epsilon^{-1} = R_\mathrm{geo}/a_\mathrm{min} = 1.8$, and an elongation $\kappa = 2.8$ (evaluated at normalised poloidal flux $\psi_\mathrm{N} = 0.95$). The toroidal magnetic field evaluated at $R = R_\mathrm{geo}$ is $B_\mathrm{geo} = 3.2$\,T, and the expected plasma current during flat-top is $I_\mathrm{p} \sim 20$\,MA. This parameter set is not too far from the set suggested for a DEMO-class ST in previous literature \cite{menard} ($R_\mathrm{geo} = 3.2$\,m, $A = 1.7$, $\kappa = 3.3$, $B_\mathrm{geo} = 2.4$\,T, $I_\mathrm{p} = 26.2$\,MA). Motivation for the choice of design parameters listed in this subsection is presented in Section~\ref{sec:limits}.

STEP will be fitted with tritium breeding blankets on the low-field side for regeneration of tritium fuel and for protection of sensitive components, such as superconducting field coils, from 14\,MeV neutrons. The total fusion power $P_\mathrm{fus}$ during flat-top operation aims for a range of 1.5 -- 1.8\,GW. A calculation example of the full steady-state power cycle for STEP, including estimated electrical power requirements for core systems, wall-plug efficiencies of auxiliary heating and current drive systems, power losses in all stages of the cycle, and heat to electricity conversion efficiency, demonstrates that a scientific fusion energy gain factor $Q_\mathrm{sci} \gtrsim 11$ ($Q_\mathrm{sci} = P_\mathrm{fus}/P_\mathrm{aux}$ where $P_\mathrm{fus}$ is the total fusion power, and $P_\mathrm{aux}$ is the auxiliary heating power injected into the plasma) is consistent with a net electricity output $P_\mathrm{net,elec.} = P_\mathrm{out} - P_\mathrm{in} > 100$\,MW~\cite{muldrew}. However, there are large uncertainties in the specific power requirements, efficiencies and losses in present stages of the STEP design, meaning that the recommended limit $Q_\mathrm{sci} \gtrsim 11$ might change with later designs. Further details on the limits for fusion power performance are given in Section~\ref{sec:fus}.

The central column has a diameter close to 3.0\,m. In order to provide space for superconducting toroidal field coils sufficiently protected from radiation and mechanical stress, only a low-capacity central solenoid can be fitted ($I_\mathrm{ind} \sim 2$\,MA), which is used for assisting start-up of scenarios only. Flat-top operation is fully non-inductive, with current driven by a combination of auxiliary systems and the bootstrap current. Two different auxiliary heating and current drive scenario options are being assessed for the current machine design. The first option includes microwave heating and current drive with both electron cyclotron (EC) and electron Bernstein wave (EBW) systems. Since the theory and modelling of EBW has less experimental validation compared to EC, a second option that includes EC systems only is being developed. These methods for auxiliary heating and current drive in STEP have been suggested after careful consideration of alternative methods~\cite{mhenderson}. Although both EC, EBW and fusion alphas primarily heat electrons, sufficient ion temperatures are still expected due to low ion heat diffusivity and heat flux predicted by gyrokinetic modelling~\cite{kennedy,giacomin} and experience from spherical tokamak experiments~\cite{kaye}. Furthermore, for larger device sizes, the confinement time is expected to increase relative to the energy exchange time, in effect leading to equilibration of electron and ion temperatures~\cite{angioni}.

Exhaust management is one of the major challenges of reactor relevant spherical tokamaks. The currently suggested design of the STEP divertor is double null~\cite{hudoba1,hudoba2,osawa}. Precise control of the separation between the two separatrices is required for sufficient distribution of the power loads between the divertors. To minimise erosion, mitigate particle flux, and buffer small transients in power, a detached plasma is needed during all main stages of operation (ramp-up, flat-top and ramp-down), which in turn requires high divertor neutral pressure, high radiation, and low $P_\mathrm{sep}/R_\mathrm{geo}$~\cite{kallenbach,henderson}. High neutral pressure is accessed by puffing D$_2$ and T$_2$ gas\footnote{While it is not impossible to design a fuel cycle that can maintain target D/T ratio with D$_2$ puffing only, the added complexity and cost to such a system makes the inclusion of T$_2$ gas preferable.} from the divertor regions together with divertor design and tuning of the pumping speed, whereas low $P_\mathrm{sep}/R_\mathrm{geo}$ is achieved by enhancing the core radiation up to 70\,\% of the total heating power using xenon-doped pellets. The materials of the high heat flux handling components remain to be confirmed, with some potential options being tungsten or tungsten alloys, or alternatively liquid tin armour.

\subsection{Integrated modelling workflow}

\begin{figure}
    \includegraphics[width=\textwidth]{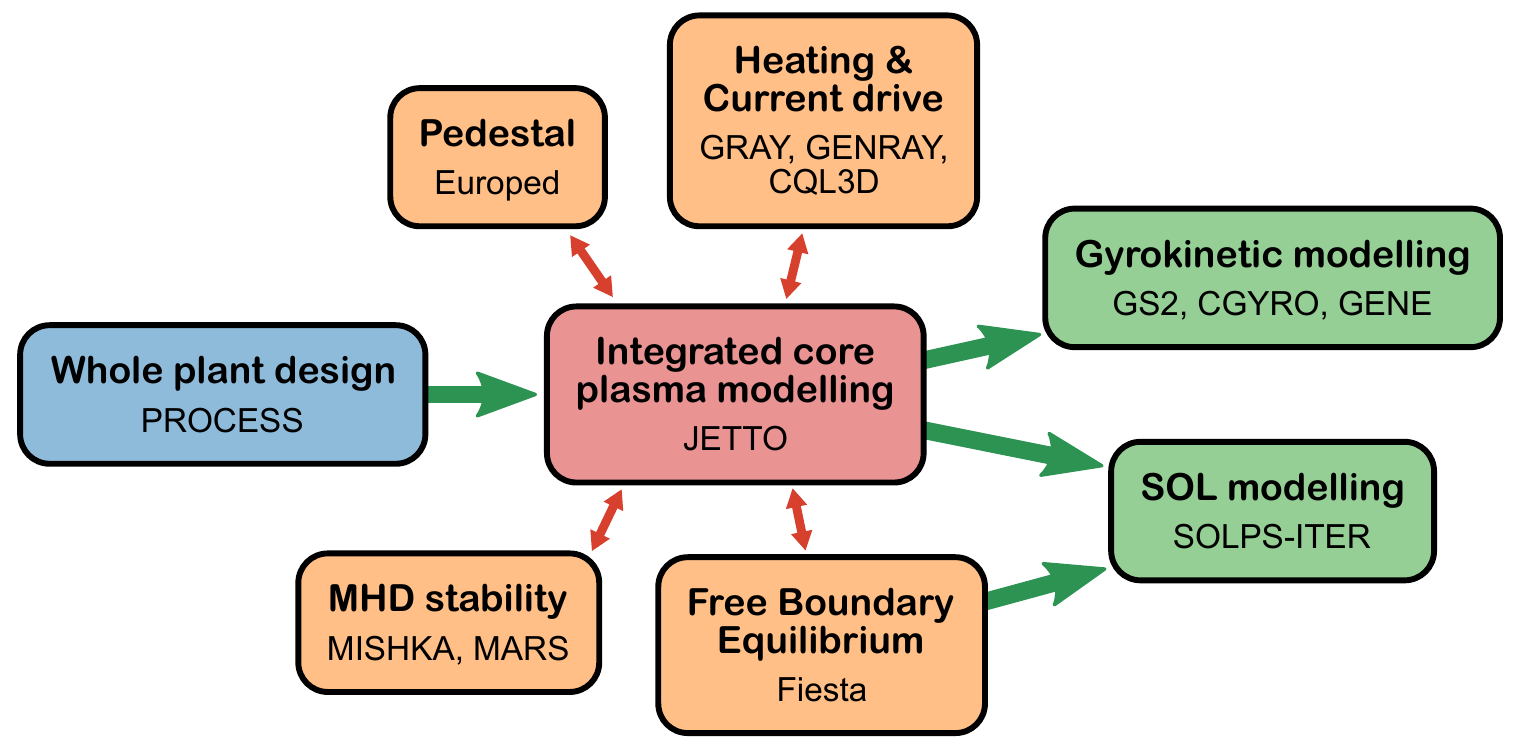}
    \caption{Diagram showing the stages of the workflow for developing flat-top operation points.}
    \label{fig:workflow}
\end{figure}

The integrated modelling workflow for developing flat-top operation points is shown in Fig.~\ref{fig:workflow}. The presented workflow is part of a larger scale iterative design process for developing more detailed engineering specifications, full scenarios including ramp-up and ramp-down, and assessments concerning e.g. cost, lifetime, environmental impact and safety of the whole scientific project~\cite{muldrew}. Several potential concepts have been developed with 0.5D systems code PROCESS~\cite{process,process2,process3}. The 1.5D integrated core plasma model JETTO~\cite{jetto} has then been used to analyse the viability of each concept from a core plasma perspective. Additional codes for MHD stability (MISHKA~\cite{mishka1,mishka2,mishka3,mishka4} and MARS~\cite{mars}), free boundary equilibrium solving (Fiesta~\cite{fiesta}), heating \& current drive modelling (GRAY~\cite{gray}, GENRAY~\cite{genray} and CQL3D~\cite{cql3d}), and pedestal modelling (Europed~\cite{europed}) are iterated against JETTO solutions to further test viability of scenarios and to gradually refine modelling assumptions against detailed physics and engineering considerations. In the final step of the flat-top modelling workflow, the concept is further analysed with gyrokinetic codes (GS2~\cite{gs2}, CGYRO~\cite{cgyro} and GENE~\cite{gene}) to verify transport assumptions, and scrape-off layer modelling (SOLPS-ITER~\cite{solps}) to study exhaust challenges.

This paper primarily presents the JETTO part of the modelling workflow used to guide the definition of the STEP operational space during flat-top. The way JETTO has been run in the presented modelling is referred to as ``assumption integration'' mode, which means that confinement is assumed rather than predicted by first-principle transport models. An empirical transport model is used to define the profile shape of the heat and particle transport coefficients. An overall scaling factor on the diffusivities are then adjusted in a feedback loop to match a target value for the total (termal + fast) normalised beta $\beta_\mathrm{N}$ while keeping the relative electron to ion heat diffusivity ratio at a fixed level indicated by results from gyrokinetic modelling. The level of confinement that results from the rescaling of the transport has to be assessed separately comparing to e.g. empirical confinement scalings. More details on the ``assumption integration'' mode for JETTO is presented in Section~\ref{sec:transp}. As theoretical understanding of transport in STEP relevant regimes increases, JETTO modelling will transition to fully predictive transport and confinement. 

The flat-top operational space is explored by performing a set of scans in key plasma parameters, with each scan point compared against operational limits that are motivated by performance requirements, engineering limitations and viability of modelling assumptions. The motivation for each of the considered operational limits in the parameter scans are presented in Section~\ref{sec:limits}. The specifics of JETTO and the assumptions used in the presented modelling are given in Section~\ref{sec:jetto}. The section also presents the two reference JETTO simulations, with each scan using one of these references as a starting point. The reference simulations, referred to as ``scan templates'', are either based on EC auxiliary heating and current drive only, or a combination of EC and EBW. Section~\ref{sec:scans} presents and discusses the results of the parameter scans. For each of the two auxiliary heating and current drive system scenarios, two candidate flat-top points have been developed based on different confinement assumptions. A higher acceptable level of confinement allows access to a broader operational space in density, temperature and auxiliary power. Two of the candidate operating points exploit these broadened parameter spaces, whereas the other two use more strict confinement assumptions. The four candidate operation points are presented in Section~\ref{sec:scen}. They have a higher degree of self-consistency compared to the simulations of the scans, including prediction of impurity density and radiation from transport and atomic physics, and $q$-profiles optimised for MHD stability. Finally, Section~\ref{sec:conc} outlines the main conclusions.

\section{Assumptions and constraints for the STEP flat-top}\label{sec:limits}
\subsection{Geometry and magnetic field}
The fusion power per unit volume of a magnetically confined D-T plasma scales as $P_\mathrm{fus}/V \propto n^2\langle\sigma v\rangle$ (any species indices are dropped in these derivations, assuming that $n_\mathrm{D} = n_\mathrm{T} = 0.5n_\mathrm{e}$ and $T_\mathrm{i} = T_\mathrm{e}$ for simplicity). For ion temperatures in the range 8 -- 30\,keV, the reactivity $\langle\sigma v\rangle \sim T^2$~\cite{menard2}, yielding $P_\mathrm{fus}/V \propto p^2 \propto \beta^2 B^4$. Furthermore, it can be shown that $\beta \propto \beta_\mathrm{N}^2(1+\kappa^2)/(f_\mathrm{BS}\sqrt{A})$, based on the assumptions that $f_\mathrm{BS} \propto a^2 p(1+\kappa^2)/(I_\mathrm{p}^2\sqrt{A}) \propto (a B/I_\mathrm{p})^2\beta(1+\kappa^2)/\sqrt{A}$~\cite{peeters}, and $a B/I_\mathrm{p} = \beta_\mathrm{N}/\beta$. This results in $P_\mathrm{fus}/V \propto \beta_\mathrm{N}^4 B^4 (1+\kappa^2)^2/(A f_\mathrm{BS}^2)$. For a spherical tokamak, it can be assumed that the elongation $\kappa^2 \gg 1$, and the bootstrap current fraction $f_\mathrm{BS} \sim 1$. In these limits, the fusion power per unit volume scales as $P_\mathrm{fus}/V \propto (\beta_\mathrm{N} B \kappa)^4/A$. Multiplying by the volume $V \propto R_\mathrm{geo}^3 \kappa/A^2$ yields the total fusion power $P_\mathrm{fus} \propto \beta_\mathrm{N}^4 B^4 \kappa^5 R_\mathrm{geo}^3/A^3$, meaning that a high normalised beta, large major radius, high elongation, strong magnetic field, and low aspect ratio are all highly beneficial for fusion power performance. 

\begin{figure}\centering
\includegraphics[width=.6\textwidth]{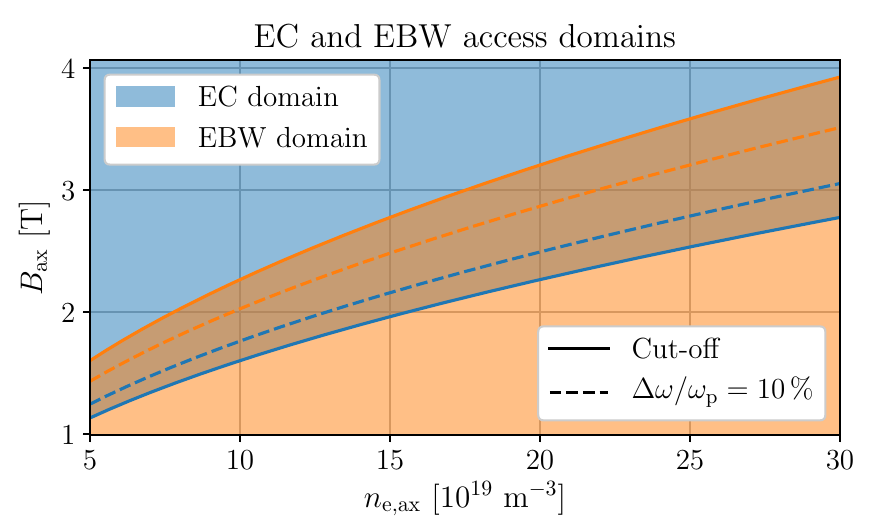}
\caption{Approximate domains in $n_\mathrm{e,ax}$,$B_\mathrm{ax}$-space (subscript ``ax'' corresponds to the value at the magnetic axis) with access to EC 2$^\mathrm{nd}$ harmonic O-mode and EBW fundamental lower X-mode. Solid curves show the corresponding cut-off boundaries, and dashed lines show the boundaries at 10\,\% frequency separation of the electron cyclotron frequency times the harmonic order relative to the plasma frequency. Details are given in Section~\ref{sec:ecebw}.}
\label{fig:ecebw}
\end{figure}

The destabilisation of resistive wall modes sets an upper limit to $\beta_\mathrm{N}$, which is discussed in more detail in Section~\ref{sec:mhd}. Depending on the length of the confinement time, different levels of auxiliary heating and current drive are required to maintain the target $\beta_\mathrm{N}$, which means that the fusion $Q$ might be too low in case the confinement is poor. This is also an aspect to take into account when setting the target $\beta_\mathrm{N}$. The elongation is limited by vertical stability constraints, although a lower internal inductance can increase the maximum elongation for vertical stability to some extent, with $\kappa_\mathrm{max} \sim 3.4 - l_\mathrm{i}$ \cite{menard2}. The magnetic field strength is primarily constrained by regimes for efficient auxiliary heating and current drive schemes, with a narrow range where both 2$^\mathrm{nd}$ harmonic O-mode (EC) and the fundamental lower X-mode (EBW) are accessible (see Fig.~\ref{fig:ecebw}). Further details are given in Sec.~\ref{sec:ecebw}. The lower limit of the aspect ratio is primarily set by the required width of the central column to accommodate toroidal field coils and a low-capacity solenoid, sufficiently shielded from particle and electromagnetic radiation. Superconducting field coils are required to reach the target toroidal field without excessive power losses due to conductor resistivity. The main limitation for the major radius is cost. An overall larger major radius can help reduce the requirements on other parameters, e.g. $\beta_\mathrm{N}$, elongation and aspect ratio, at maintained fusion power performance. However, with the restriction on the magnetic field for EC and EBW access, the rod current has to grow proportionally with the major radius, meaning that the central column needs to grow with increased capacity requirement of the toroidal field coils. Balancing the major radius against performance and cost, the presently suggested design point is at $R_\mathrm{geo} = 3.6$\,m. This choice is further motivated in~\cite{muldrew}. The width requirement on the central column at that size suggests a minor radius of 2.0\,m, i.e. an aspect ratio $A = 1.8$.

\subsection{Fusion power and fusion $Q$}\label{sec:fus}
As already mentioned in Section~\ref{sec:intro}, the target fusion power is in the range 1.5 -- 1.8 GW, with a $Q_\mathrm{sci} > 11$. Even in the case of optimal $Q_\mathrm{sci}$, where alpha heating alone is sufficient to maintain thermonuclear ion temperatures and auxilliary systems operate at minimal power for shaping the current profile, at least 1.5\,GW of fusion power might be required to balance power requirements from systems such as field coils and cooling, considering the expected efficiency of energy conversion systems from the energetic neutrons to electric power. A fusion power of 1.8\,GW corresponds to a neutron production rate of $6.4\times 10^{20}$\,s$^{-1}$, and to an average power flux of 2.7\,MW/m$^2$ carried by neutrons as they cross the separatrix. Assuming for simplicity that all fusion reactions originate from the magnetic axis, the local power flux carried by neutrons would be around 5\,MW/m$^2$ on the first wall at the outer mid-plane. 

\subsection{EC and EBW access}\label{sec:ecebw}
Models for EC heating and current drive have been validated against several ex\-pe\-ri\-ments, and the underlying theory is well understood. This contrasts to the lower level of maturity for EBW heating and current drive prediction. However, EBW remains a promising candidate for non-inductive current drive, due to the expected efficiency of power deposition to the high-energy tail of the electron distribution. In addition, a large Ohkawa current is predicted with the high trapped electron fraction in spherical tokamak regimes~\cite{ebwst1,ebwst2}, which potentially makes EBW current drive significantly more efficient than ECCD for STEP. Two scenarios are being considered in parallel using EC heating and current drive only and using a combination of EC and EBW. The EC only scenario is modelled with a higher degree of confidence, due to the extensive validation of EC heating and current drive models against experiments, whereas simulations of the EC+EBW scenario suggest a more efficient power plant, requiring lower recirculation power~\cite{muldrew}.

ECCD has poor off-axis current drive efficiency due to the large trapped electron fraction. It is used in both STEP scenarios for on-axis auxiliary current drive, which is required to prevent current hole formation, with the bootstrap current vanishing on-axis (motivation for why on-axis current holes should be avoided is outlined in Section~\ref{sec:mhd}). On the other hand, EBW is predicted to have low current drive efficiency on-axis, since the Doppler shift of the cyclotron harmonics causes all power to be deposited off-axis ($\rho_\mathrm{tor} \gtrsim 0.5$) before propagating to the core. Scans have been conducted in injector locations, steering mirror angles, and beam frequencies during STEP flat-top operation~\cite{freethy}, using GRAY~\cite{gray} benchmarked against CQL3D~\cite{cql3d} for EC modelling, and GENRAY~\cite{genray}/CQL3D for EBW modelling. It was shown that optimal ECCD efficiency is obtained by coupling to the 2$^\mathrm{nd}$ harmonic O-mode. EBW heating and current drive were shown to be the most feasible by access to the fundamental X-mode via conversion from O-mode. The X-mode wave then naturally transforms to a Bernstein wave (so called O-X-B coupling~\cite{oxb}). 

A simple access criterion for EC heating and current drive is that the density cut-off is not exceeded on-axis, where the density is expected to be the highest. For the $m^\mathrm{th}$ harmonic O-mode, this criterion is
\begin{equation}\label{eq:ecco}
    \omega_\mathrm{c,e,ax} > \frac{\omega_\mathrm{p,ax}}{m}~\Leftrightarrow~B_\mathrm{ax} > \sqrt{\frac{n_\mathrm{e,ax}m_\mathrm{e}}{m^2\epsilon_0}},
\end{equation}
which corresponds to the blue domain in Fig.~\ref{fig:ecebw} for $m = 2$. This simple access criterion does not include effects, like harmonic shadowing, that arise from variation of density and magnetic field across the plasma. Nevertheless, it gives an indication of EC access that can be used for analysing the operational domains within the parameter scans in Section~\ref{sec:scans}. 

A corresponding criterion can be formulated for EBW access. When the right hand cut-off frequency exceeds the frequency of the launched X-mode, a cut-off and the adjacent upper hybrid layers are guaranteed to exist in the plasma. The latter becomes the mode conversion layer for conversion into the EBW wave if warm plasma effects are accounted for. For an arbitrary harmonic $m \geq 1$, the criterion for the cut-off frequency to exceed the X-mode frequency can be formulated as
\begin{equation}\label{eq:ebco}
    \frac{1}{2}\left(\sqrt{\omega_\mathrm{c,e,ax}^2+4\omega_\mathrm{p,ax}^2} - \omega_\mathrm{c,e,ax}\right) > m\omega_\mathrm{c,e,ax}~\Leftrightarrow~\omega_\mathrm{c,e,ax} < \frac{\omega_\mathrm{p,ax}}{\sqrt{m(m+1)}}.
\end{equation}
Evaluated at $m = 1$ for the fundamental resonance, the inequality of eq.~(\ref{eq:ebco}) corresponds to the orange domain in Fig.~\ref{fig:ecebw}.

\subsection{MHD instabilities and $q$-profile shaping}\label{sec:mhd}
Important MHD instabilities to consider in a spherical tokamak include ballooning in\-sta\-bi\-li\-ties, sawteeth, neoclassical tearing modes (NTMs), resistive wall modes (RWMs), Alfvén eigenmodes (AEs), and edge-localised modes (ELMs). Many MHD in\-sta\-bi\-li\-ties can be avoided by strategic shaping of the $q$-profile. On-axis current hole formation with an associated strongly reversed shear $q$-profile should generally be avoided, since it can result in internal transport barriers that lead to disruptions~\cite{challis}. Disruption events can be catastrophic at the plasma currents foreseen for STEP (see details in Section~\ref{sec:cur}). Strongly reversed $q$-profiles also risk destabilisation of double tearing modes~\cite{dtm1,dtm2} and other mixed modes~\cite{mm1}, and it can destabilise AEs~\cite{ae1,ae2}. Non-inductive, high bootstrap current experiments such as STEP are vulnerable to current hole formation, since the bootstrap current density vanishes on-axis (the pressure gradient vanishes on-axis, and the bootstrap current contribution from potato orbits is expected to be negligible~\cite{helander}), and there is no in\-duc\-tive current to fill the resulting gap. Thus, on-axis auxiliary current drive, e.g. from ECCD, is essential to complement the bootstrap current and to avoid the onset of a current hole. 

The current density should be sufficiently low to avoid current driven MHD instabilities, such as sawteeth and low order NTMs occuring at $q = 2$ (which are likely to cause disruptions). Thus, $q_\mathrm{min}$ should be at least 2, which is consistent with direct access to second ballooning mode stability~\cite{mercier}. For NTMs, the strong curvature stabilisation at low aspect ratio~\cite{haye} compared to the destabilising bootstrap term gives stability to $n = 1$ and $n = 2$ instabilities. In the shaping of the ECCD profile, the aim has been to achieve a relatively broad region around the axis ($\rho_\mathrm{tor} \lesssim 0.4$) with a low core magnetic shear out to $q \sim 3$. Since the equilibria tend to have low shear in the core, it is found that $q_\mathrm{min}$ above about 2.2 is necessary to avoid infernal modes~\cite{manickam} being unstable, or close to unstable, and coupling to RWMs. Tools have been developed for $q$-profile optimisation with ECCD shaping in STEP scenarios based on a genetic algorithm~\cite{marsden} and multi-objective Bayesian optimisation~\cite{brown}.

There is an upper limit of $\beta_\mathrm{N}$ ($\beta_\mathrm{N}$ includes both thermal and fast particle pressure, e.g.\ from alpha particles) for the destabilisation of RWMs, which grow on time scales similar to the vertical field diffusion time, a.k.a.\ the wall time. The degree of de\-sta\-bi\-li\-sa\-tion can be quantified by the parameter
\begin{equation}\label{eq:cbdef}
    C_\beta = \frac{\beta_\mathrm{N} - \beta_\mathrm{N}^\mathrm{no{\mhyphen}wall}}{\beta_\mathrm{N}^\mathrm{ideal{\mhyphen}wall} - \beta_\mathrm{N}^\mathrm{no{\mhyphen}wall}},
\end{equation}
where $\beta_\mathrm{N}^\mathrm{no{\mhyphen}wall}$ and $\beta_\mathrm{N}^\mathrm{ideal{\mhyphen}wall}$ are the RWM stability limits without and with a perfectly conducting wall, respectively. Operation at $C_\beta \gtrsim 1$, i.e. $\beta_\mathrm{N} \gtrsim \beta_\mathrm{N}^\mathrm{ideal{\mhyphen}wall}$, will result in uncontrollable growth of RWMs. On the other hand, at domains of weakly positive $C_\beta$, with $\beta_\mathrm{N} \approx \beta_\mathrm{N}^\mathrm{no{\mhyphen}wall}$, RWMs are likely to be stabilised by passive stabilisation mechanisms, such as plasma rotation~\cite{rot} or kinetic resonances. At slightly higher $C_\beta \lesssim 0.5$, RWMs can be actively mitigated using RMP coils (RMP = Resonant Magnetic Perturbation). Eq.~(\ref{eq:cbdef}) considers destabilisation of the $n = 1$ toroidal mode, which has the lowest ideal-wall limit and is the most likely RWM to trigger disruption events. Multiple toroidal mode numbers are potentially destabilised at $\beta_\mathrm{N}$ above the corresponding no-wall limits, and need to be considered in the design of the active RWM mitigation systems.

Experiments have shown that the upper $\beta_\mathrm{N}$ limit can be expressed as a function of the pressure peaking $F_p = p_\mathrm{ax}/\langle p\rangle$ according to $0.2 + 12.5/F_p$~\cite{nsrw}. Assuming that this corresponds to the $n = 1$ ideal-wall limit, $C_\beta = 1$ matches this parametrisation. Simulations with the MHD code MARS~\cite{mars} have shown that the no-wall limit is more or less independent of the pressure peaking, with a value close to 3.7 for STEP-like flat-tops. In the scans of Section~\ref{sec:scans}, an approximation of the destabilisation parameter
\begin{equation}\label{eq:cbeta}
    C_\beta \approx \frac{(\beta_\mathrm{N} - 3.7)F_p}{12.5 - 3.5 F_p}
\end{equation}
has been evaluated at each scan point based on the above assumptions, with an operational regime estimated as $C_\beta \leq 0.5$. This limit should only be interpreted as indicative of RWM stability, since there is additional physics that influences no-wall and ideal-wall $\beta_\mathrm{N}$ limits, which is not captured by the simple relationship of eq.~(\ref{eq:cbeta}). The candidate flat-top points presented in Section~\ref{sec:scen} have been further analysed with MARS to get a more accurate prediction of $C_\beta$.

Regarding stabilisation of RWMs by rotation, STEP does not have NBI systems, which would otherwise be the primary source of toroidal momentum. Although the alpha particles are born isotropic, orbit effects during slowing down induce a net torque ($\bm{J}\times\bm{B}$-driven and collisional), which can be enhanced by toroidal field ripple and RMPs. Since the momentum diffusivity is not known for STEP relevant regimes, there are large uncertainties in the expected rotation resulting from the induced alpha particle torque. Consequently, rotation has not been included in our current JETTO calculations. 

The Greenwald density $n_\mathrm{GW}[10^{20}\,\mathrm{m}^{-3}] = I_\mathrm{p}[\mathrm{MA}]/(\pi a[\textrm{m}]^2)$ is an empirical limit to consider for operational stability. Experiments have been able to demonstrate core density operation far above the Greenwald limit in scenarios with benign ELM activity~\cite{ngw}. This indicates that the Greenwald density is not a strict limit of the core plasma, but rather a limit more closely associated with pedestal or SOL behaviour. In the parameter scans of Section~\ref{sec:scans}, line averaged core densities are allowed to exceed the Greenwald limit by up to 20\,\%. ELM stability is analysed separately for individual candidate flat-top operating points. 

\subsection{Plasma current and internal inductance}\label{sec:cur}
A high plasma current is beneficial for fusion power performance, since it allows access to higher densities and kinetic pressures at constant Greenwald fraction and $\beta_\mathrm{N}$. However, the forces on surrounding vessel structures during disruption events scale as $I_\mathrm{p}^2$, ultimately setting an upper limit to the safe operation of the plasma current. Other risks with high plasma currents also include runaway electrons, with a generation rate scaling exponentially with $I_\mathrm{p}$, and current driven global MHD instabilities. The agreed upper limit of $I_\mathrm{p} = 25$\,MA is based on current (limited) understanding of the ability to manage the above-mentioned risks, which means that the limit is likely to change in more mature stages of the design. In the scans of Section~\ref{sec:scans}, a slightly more conservative limit of $I_\mathrm{p} = 23$\,MA has been chosen. This is because $I_\mathrm{p} < 25$\,MA is satisfied in almost the entire scan spaces, whereas $I_\mathrm{p} < 23$\,MA reveals domains of high current in multi-dimensional presentations of the scan space (see e.g. Fig.~\ref{fig:scan1}.c).

A broad current density profile, resulting in a low internal inductance, is beneficial for vertical stability. The current density profile can be directly controlled using auxiliary current drive systems, and indirectly via the bootstrap current, which is influenced e.g. by the pressure peaking. The individual scan points of Section~\ref{sec:scans} have not been checked against a vertical stability limit, since the auxiliary current drive has been extrapolated from the template simulations using simplifying assumptions (see Section~\ref{sec:hcd}), without regard for the degrees of freedom of the systems for current profile shaping. On the other hand, in the development of the candidate flat-top points of Sec.~\ref{sec:scen}, the auxiliary systems have been tuned to result in a broad current profile and low $l_\mathrm{i}$. Using the simple limit $\kappa_\mathrm{max} = 3.4 - l_\mathrm{i}$~\cite{menard2} (the normalised inductance $l_\mathrm{i}$ is defined according to eq.~(10) in~\cite{luce}), the STEP elongation $\kappa = 2.8$ would imply $l_\mathrm{i} < 0.6$ for vertical controllability. As shown in Table~\ref{tab:cand}, the achieved normalised inductance never exceeds 0.53. 

\subsection{Energy confinement}\label{sec:econf}

Turbulent transport in STEP high beta plasmas is expected to be electromagnetic and not to be well described by current state-of-the-art reduced core plasma transport models (e.g. TGLF~\cite{tglf} and QuaLiKiz~\cite{qlk}). In this paper, the modelled STEP scenarios will be guided using empirical models of the core transport. Computationally intensive first principles-based calculations of turbulent transport in STEP scenarios developed in this paper are reported in \cite{kennedy} and \cite{giacomin}. We proceed by comparing design points to empirical scalings of the confinement time in order to quantify the viability of the energy confinement implied by the transport assumptions. As a proxy, the ITER energy confinement scaling $\tau_\mathrm{IPB98(y,2)}$~\cite{iter} has been selected for assessing the confinement assumptions in this paper. Despite being based on conventional tokamaks, the scaling is developed from a relatively large database of experiments (more than 1\,300 H-mode pulses from 9 different tokamaks), meaning that its validity spans a broad plasma parameter space. The aspect ratio of the tokamaks in the database stretches between 2.5 (DIII-D) and 5.5 (PBX-M).

The confinement assumption can be quantified by the confinement factor, defined as $H_\mathrm{IPB98(y,2)} = \tau_\mathrm{E}/\tau_\mathrm{IPB98(y,2)}$. Since the confinement is uncertain and has potential for op\-ti\-mi\-sa\-tion, we seek the minimum confinement assumption that satisfies the other constraints for a non-inductive burning plasma. The scans of Section~\ref{sec:scans} check the viability of the confinement assumption with fixed values of the maximum confinement factors. Two limiting values are selected: one ``conservative'' limit of 1.2, and one ``optimistic'' limit of 1.4. It is presently unknown whether realistic values of the confinement factor in STEP relevant regimes lie between, above or below these two values, making the specific limits somewhat arbitrary at present stages of the modelling. However, the limits are sufficient to provide guidelines for target regimes within the operational spaces set by other constraints. As will be demonstrated in Sections~\ref{sec:scans} and \ref{sec:scen}, the conservative limit of 1.2 is possible to satisfy, but only with the high current drive efficiency of EBW, whereas the optimistic limit of 1.4 can also be satisfied in scenarios without EBW. 

The ITER scaling of the confinement time $\tau_\mathrm{IPB98(y,2)}$ scales with respect to net power as $P_\mathrm{net}^{-0.69}$, where $P_\mathrm{net} = P_\mathrm{tot} - \mathrm{d}W_\mathrm{th}/\mathrm{d}t$, $P_\mathrm{tot} = P_\mathrm{aux} + P_\alpha + P_\mathrm{ohm}$ and $W_\mathrm{th}$ is the (thermal) stored energy. The scaling typically does not include contribution to $P_\mathrm{net}$ from the radiated power, under the assumption that $P_\mathrm{rad} \ll P_\mathrm{tot}$. However, this assumption does not hold for STEP plasmas, since a significant fraction of seeded impurity radiation is likely to be required for detachment access ($P_\mathrm{rad}/P_\mathrm{tot} \sim 0.7$). A radiation corrected confinement factor, here referred to as $H_{98}^*$, has been derived for highly core radiated plasmas using ASTRA/TGLF~\cite{lux}. The definition replaces $P_\mathrm{net}$ with $P_\mathrm{net}^* = P_\mathrm{net} - \gamma P_\mathrm{rad}$ for some coefficient $0 < \gamma < 1$, meaning that $H_{98}^* = \tau^*_\mathrm{E}/\tau_{98}^* \propto (W_\mathrm{th}/P_\mathrm{net}^*)/(P_\mathrm{net}^*)^{-0.69} \propto (P_\mathrm{net} - \gamma P_\mathrm{rad})^{-0.31}$. A high $\gamma P_\mathrm{rad}$ implies a high $H_{98}^*$, and vice versa. That is, if $\gamma P_\mathrm{rad}$ is large, then high confinement is required in order to be consistent with a given density and temperature regime. The studies in~\cite{lux} found that radiation in the deep core influences the confinement more than radiation closer to the separatrix. More specifically, $\gamma$ can be estimated by taking 60\,\% of the radiation inside $\rho_\mathrm{tor} = 0.75$ into account. Note that this $\gamma$ was derived for the DEMO conventional tokamak, and it is presently unknown how well it extrapolates to STEP. For lack of better alternatives, it is used in the presented confinement assessments. Since the radiated power distribution is not self-consistently modelled during the parameter scans of Section~\ref{sec:scans}, $\gamma = 0.6$ is chosen, which is a conservative assumption in that it assumes all radiation to impact the confinement, and $H_{98}^*$ is maximised. On the other hand, the derived flat-top points of Section~\ref{sec:scen}, which includes more self-consistent modelling of impurities and the radiated power distribution, only consider the radiation inside $\rho_\mathrm{tor} = 0.75$ in the $P_\mathrm{net}^*$ estimate.

Several empirical scaling laws suggest a weaker density scaling compared to $\tau_\mathrm{IPB98(y,2)} \propto \langle n_\mathrm{e}\rangle^{0.41}$, such as $\tau_\mathrm{Petty08} \propto \langle n_\mathrm{e}\rangle^{0.32}$~\cite{petty}, $\tau_\mathrm{ITPA20} \propto \langle n_\mathrm{e}\rangle^{0.24}$~\cite{ITPA20}, and $\tau_\mathrm{ITPA20{\mhyphen}IL} \propto \langle n_\mathrm{e}\rangle^{0.15}$~\cite{ITPA20}. Scaling laws derived from spherical tokamak databases even suggest a weakly negative scaling with respect to density, such as $\tau_\mathrm{NSTX19} \propto \langle n_\mathrm{e}\rangle^{-0.05}$~\cite{buxton}, and $\tau_\mathrm{MAST09} \propto \langle n_\mathrm{e}\rangle^{-0.06}$~\cite{mast}. However, most scaling laws are based on a smaller database of pulses than $\tau_\mathrm{IPB98(y,2)}$, in particular the purely ST based scaling laws. Nevertheless, given the uncertainty in the density scaling of confinement in STEP relevant regimes, it is prudent to explore the optimisation of operating points using several different scalings, and to design a machine that can accommodate a range of operating points at different densities. 

\subsection{Power losses and detachment access}\label{sec:ploss}
STEP will need to operate with a detached plasma in most phases of operation to provide sufficient protection of the divertor. The peak heat loads associated with an attached divertor plasma would far exceed the expected engineering limits of $\sim 10$\,MW/m$^2$ in steady-state. The operational challenge is to integrate the detachment with the plasma core, and to make the detached plasma operation robust to small fluctuations of the plasma. Some of the present exhaust assumptions and limitations have been further addressed using SOL, wall and divertor modelling with SOLPS-ITER~\cite{osawa}. Since JETTO is a core plasma model, boundary conditions in e.g. temperature, density and impurity fraction need to be applied at the last closed flux surface, with no indication of whether the applied conditions are compatible with a divertor plasma solution. However, a reduced model of the SOL and divertor can be coupled to JETTO to provide more realistic boundary constraints, while also assessing the detachment requirements at the divertor. This can provide a useful check for operational regimes in the JETTO parameter scans of Sections~\ref{sec:scans}, since it is unfeasible to complement every single scan point with detailed exhaust analyses from more advanced, computationally heavy SOL/divertor models.

Detachment access can be estimated using a detachment qualifier
\begin{eqnarray}\label{eq:qdet}
    \fl q_\mathrm{det} = 1.3\frac{P_\mathrm{sep}[\mathrm{MW}]}{R_\mathrm{geo}[\mathrm{m}]}\frac{5\,\mathrm{mm}}{\lambda_\mathrm{int}[\mathrm{mm}]}\left(\frac{1.65\,\mathrm{m}}{R_\mathrm{geo}[\mathrm{m}]}\right)^{0.1} \left(\left[1+\sum_Z{f_Z c_{Z,\mathrm{div}}}\right]p_\mathrm{0,div}[\mathrm{Pa}]\right)^{-1},
\end{eqnarray}
where detachment is accessed at $q_\mathrm{det} < 1$ and partial detachment at $q_\mathrm{det} \approx 1$. $\lambda_\mathrm{int}$ is the SOL power width with the effect of power spreading taken into account ($\lambda_\mathrm{int} = \lambda_q + 1.64S_\mathrm{PS}$, $S_\mathrm{PS}$ is the power spreading factor~\cite{eich}). $f_Z$ is the radiation efficiency of impurity $Z$ relative to deuterium, and $c_{Z,\mathrm{div}}$ is the concentration of the impurity in the divertor region. $p_\mathrm{0,div}$ is the divertor neutral pressure. 

Argon is planned to be the primary impurity species seeded for detachment control in STEP, which is expected to have a radiation efficiency $f_\mathrm{Ar} = 90$. STEP needs to operate with the minimal amount of argon that is still sufficient to reach the required degree of detachment, since even a small amount of argon leaking into the core can substantially degrade the fusion power performance by dilution ($P_\mathrm{fus} \propto (1 - 18 c_\mathrm{Ar,core})^2$, e.g. 0.5\,\% of argon reduces the fusion power by more than 17\,\%). Reduced core ion temperature resulting from impurity radiation also risks reducing the fusion reaction cross section. The level of screening of argon from the core can be quantified with the enrichment factor $\eta_Z = (n_{Z,0,\mathrm{div}}/n_{\mathrm{D},0,\mathrm{div}})/(n_{Z,\mathrm{core}}/n_\mathrm{e,core})$ (subscript 0 refers to neutral density). Experiments at ASDEX Upgrade have demonstrated an argon enrichment factor of about 2~\cite{kallenbach2}. A higher enrichment factor is expected for STEP due to neoclassical screening in the pedestal~\cite{dux,field}. A current estimate of the maximum $c_\mathrm{Ar,div}$ without unacceptable dilution in the core is 2 -- 3\,\%.

Other parameters that play a role in the detachment access according to eq.~(\ref{eq:qdet}) are $P_\mathrm{sep}/R_\mathrm{geo}$, $\lambda_\mathrm{int}$ and $p_{0,\mathrm{div}}$. An upper limit of $P_\mathrm{sep}/R_\mathrm{geo} = 45$\,MW/m (corresponding to $P_\mathrm{sep} = 45\times 3.6 = 162$\,MW) has been chosen to define the operational domain in the scans of Section~\ref{sec:scans}. In order for the limit to correspond to $q_\mathrm{det} = 1$ in eq.~(\ref{eq:qdet}), it is assumed that $(1 + 90 c_\mathrm{Ar,div})\lambda_\mathrm{int}p_{0,\mathrm{div}} \approx 270$\,mmPa. An example of a combination of assumptions compatible with this number is $c_\mathrm{Ar,div} = 2.9$\,\%, $\lambda_\mathrm{int} = 5.0$\,mm, and $p_{0,\mathrm{div}} = 15$\,Pa. These are all values believed to be compatible with core and edge operational constraints under current assumptions. However, $\lambda_\mathrm{int}$, and $p_\mathrm{0,div}$ are challenging to estimate in the present stages of modelling, requiring more self-consistent pump, SOL, impurity seeding and transport, and overall core confinement modelling to narrow down uncertainties. Later iterations of the divertor design are also likely to shift these parameter values. Despite inherent inaccuracies of the correspondence between the $P_\mathrm{sep}/R_\mathrm{geo} < 45$\,MW/m limit and the domain of detachment access, it remains a useful guideline, since $P_\mathrm{sep}$ is straightforward to evaluate in the present stages of core plasma modelling.

\subsection{Summary of parameters and their constraints}\label{sec:limsum}
The operational limits that are considered for all of the scans are the following:
\begin{enumerate}
    \item $1.5\,\textrm{GW} \leq P_\mathrm{fus} \leq 1.8\,\textrm{GW}$,
    \item $Q_\mathrm{sci} \geq 11$,
    \item $I_\mathrm{p} \leq 23\,\textrm{MA}$,
    \item $C_\beta \leq 0.5$,
    \item $P_\mathrm{sep}/R_\mathrm{geo} \leq 45\,\textrm{MW/m}$,
    \item $H_\mathrm{98}^* \leq 1.2$ (conservative) or $H_\mathrm{98}^* \leq 1.4$ (optimistic).
\end{enumerate}
In addition, access to EC 2$^\mathrm{nd}$ harmonic O-mode and EBW fundamental lower X-mode coupling is considered, according to the cut-off domains estimated in Section~\ref{sec:ecebw}.

\section{Integrated core plasma modelling with JETTO}\label{sec:jetto}

To examine steady-state flat-top scenarios, we use JETTO in assumption integration mode, in which confinement is assumed. In this way of running JETTO, anomalous heat and particle transport is not predicted from turbulent physics. Rather, a target regime in density and temperature is set, and the anomalous transport is artificially rescaled until the target is reached. Section~\ref{sec:transp} presents this approach and its motivation in more detail. The main model inputs are:
\begin{itemize}
    \item Size and shape of the separatrix
    \item Densities and temperatures at the separatrix
    \item Vacuum toroidal magnetic field at $R = R_\mathrm{geo}$: $B_\mathrm{geo}$
    \item Normalised beta: $\beta_\mathrm{N} = \beta_\mathrm{N,th} + \beta_\mathrm{N,fast}$
    \item Relative electron to ion anomalous heat diffusivity
    \item Greenwald density fraction, $f_\mathrm{GW} = \langle n_\mathrm{e}\rangle/n_\mathrm{GW}$
    \item Auxiliary power densities: $q_\mathrm{EC}$ and $q_\mathrm{EBW}$ [W/m$^3$]
    \item EC current drive efficiency: $\zeta_\mathrm{CD}$,
    \item EBW current density: $J_\mathrm{EBW}$
\end{itemize}
Some of the main outputs are:
\begin{itemize}
    \item Plasma current density: $J_\mathrm{p}$ [A/m$^2$]
    \begin{itemize}
        \item Predicted from bootstrap current $J_\mathrm{BS}$, EC current drive $J_\mathrm{EC}$ and (prescribed) EBW current $J_\mathrm{EBW}$
    \end{itemize}
    \item Heat and particle transport
    \begin{itemize}
        \item Neoclassical is predicted, and anomalous is artificially rescaled
    \end{itemize}
    \item Kinetic profiles (densities \& temperatures)
    \item Fusion power density: $q_\mathrm{fus}$
    \item Alpha heating: $q_{\alpha,\mathrm{e}}$ and $q_{\alpha,\mathrm{i}}$
    \item 2D magnetic equilibrium: $B_R(R,z)$, $B_z(R,z)$ and $B_\mathrm{tor}(R,z)$
    \item Power flux across the separatrix: $P_\mathrm{sep}$
\end{itemize}
There are some differences between the modelling assumptions of the parameter scans of Sec.~\ref{sec:scans} and the list of candidate flat-top operating points in Sec.~\ref{sec:scen}, which impacts some of the set of inputs and outputs. For instance, the radiation fraction $f_\mathrm{rad} = P_\mathrm{rad}/(P_\mathrm{aux} + P_\alpha + P_\mathrm{ohm})$ and the effective charge $Z_\mathrm{eff} = \sum_i Z_i^2 n_i/\sum_i Z_i n_i$ are inputs in the scans, but they are outputs in the candidate flat-top calculation. Details on this is outlined in Sec.~\ref{sec:imprad}.

\subsection{Scan templates}\label{sec:templates}

\begin{figure}
    \centering
    \includegraphics[width=\textwidth]{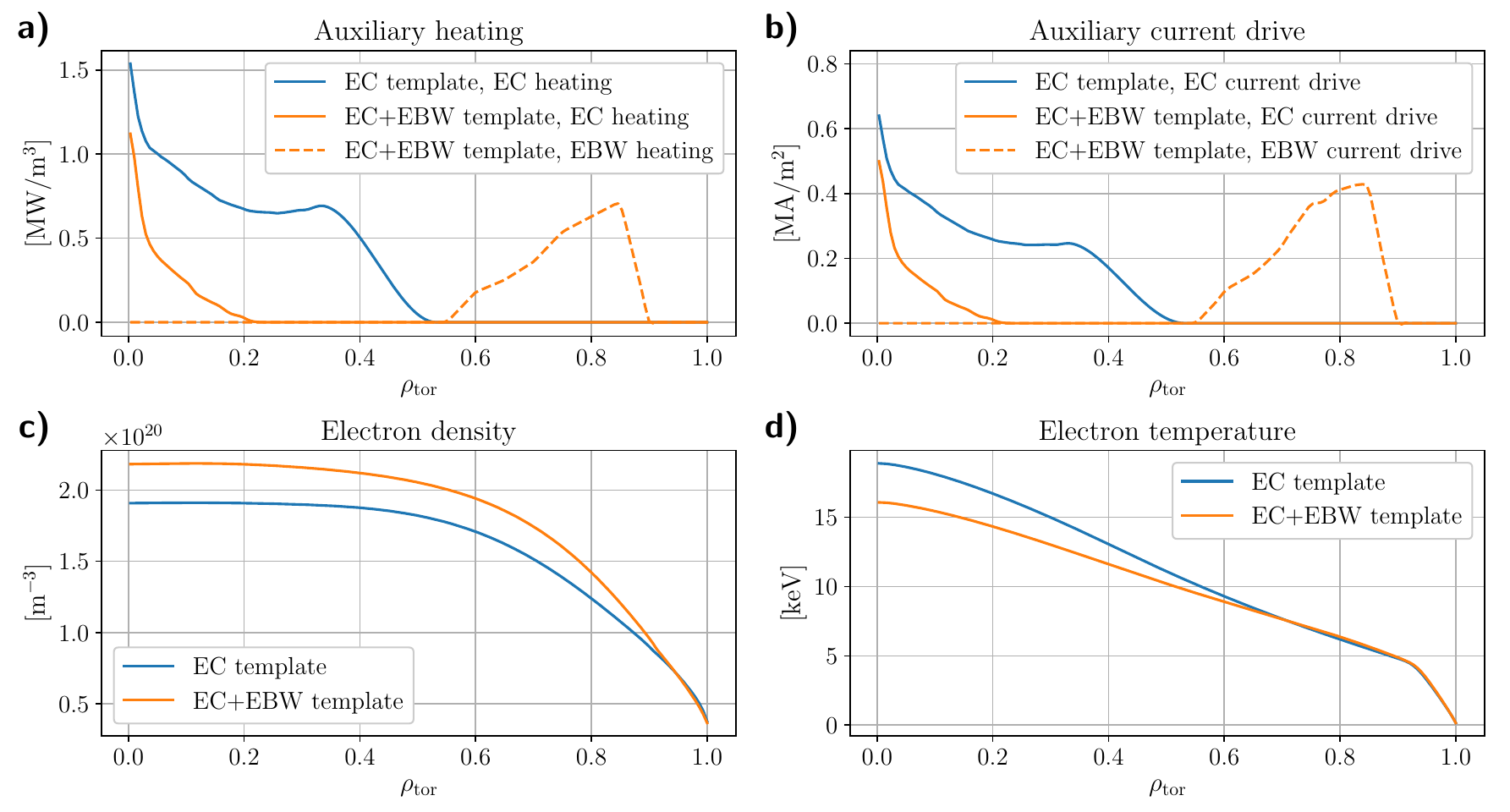}
    \caption{Top figures: Auxiliary heating and current drive for the two template simulations used in the scans. Bottom figures: Electron density and temperature for the template simulations.}
    \label{fig:eej}
\end{figure}
\begin{figure}
    \centering
    \includegraphics[width=\textwidth]{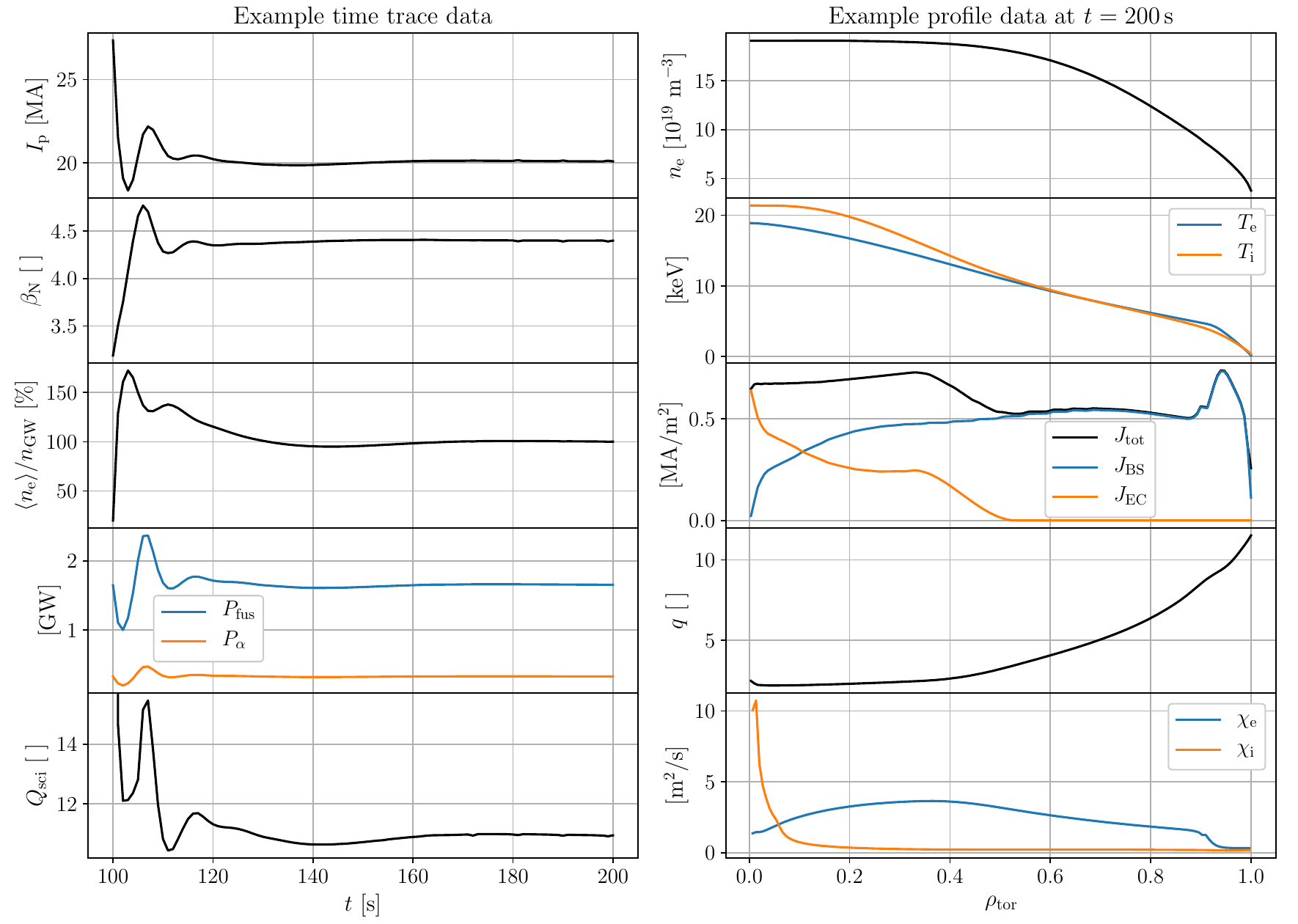}
    \caption{Selection of JETTO outputs from the EC template simulation. $\rho_\mathrm{tor}$ is the square root of the normalised toroidal magnetic flux.}
    \label{fig:jetto}
\end{figure}
\begin{table}
    \centering
    \caption{Summary of 0D outputs from the two template simulations.\vspace{2mm}}
    \begin{tabular}{rlcc}\hline
        & & \textbf{EC}~\cite{ectemp} & \textbf{EC+EBW}~\cite{ebtemp} \\\hline
        $P_\mathrm{fus}$ & [GW] & 1.65 & 1.77 \\
        $Q_\mathrm{sci}$ & [ ] & 10.94 & 13.60 \\
        $P_\mathrm{EC}$ & [MW] & 150.00 & 9.00 \\
        $P_\mathrm{EBW}$ & [MW] & --- & 120.00 \\
        $\langle n_\mathrm{e}\rangle$ & [10$^{19}$ m$^{-3}$] & 15.40 & 17.35 \\
        $\langle n_\mathrm{e}\rangle_\mathrm{line}/n_\mathrm{GW}$ & [\%] & 100.13 & 100.02 \\
        $\langle T_\mathrm{e}\rangle$ & [keV] & 9.81 & 9.05 \\
        $\langle T_\mathrm{i}\rangle/\langle T_\mathrm{e}\rangle$ & [ ] & 1.07 & 1.04 \\
        $I_\mathrm{p}$ & [MA] & 20.10 & 22.88 \\
        $I_\mathrm{BS}/I_\mathrm{p}$ & [\%] & 87.88 & 81.74 \\
        $I_\mathrm{EC}$ & [MA] & 2.17 & 0.15 \\
        $I_\mathrm{EBW}$ & [MA] & --- & 3.95 \\
        $I_\mathrm{EC}/P_\mathrm{EC}$ & [kA/MW] & 14.48 & 16.14 \\
        $I_\mathrm{EBW}/P_\mathrm{EBW}$ & [kA/MW] & --- & 32.89 \\
        $\beta_\mathrm{N}$ & [ ] & 4.40 & 3.93 \\
        $B_\mathrm{axis}$ & [T] & 2.59 & 2.57 \\
        $l_\mathrm{i}$ & [ ] & 0.50 & 0.39 \\
        $l_\mathrm{i}(3)$ & [ ] & 0.28 & 0.22 \\
        $q_\mathrm{min}$ & [ ] & 2.18 & 3.14 \\
        $q_{95}$ & [ ] & 9.35 & 8.00 \\
        $P_\mathrm{rad}$ & [MW] & 336.02 & 336.72 \\\hline
    \end{tabular}
    \label{tab:templates}
\end{table}

The parameter scans of Sec.~\ref{sec:scans} are done by varying key plasma parameters relative to a given reference case, here referred to as a scan template. Two different scan templates are used, which are based on the two different auxiliary heating and current drive schemes that are suggested for STEP. Consequently, these are referred to as the EC template~\cite{ectemp} and the EC+EBW template~\cite{ebtemp}. A summary of 0D data for the two templates is presented in Table~\ref{tab:templates}. The auxiliary heating power and current drive profiles for the two scan templates, together with electron densities and temperatures, are shown in Fig.~\ref{fig:eej}. 

All of the simulations presented in this paper target single stationary flat-top operating points, rather than dynamical scenarios. Each simulation is continued until outputs have sufficiently converged. Since current diffusion time scales are of the order 10$^3$\,s in most of the plasma, the plasma resistivity in the current diffusion equation has been artificially scaled up by a constant factor. This scaling should not impact the outcome of converged solution. A set of example outputs from a JETTO simulation is presented in Fig.~\ref{fig:jetto} (the particular simulation presented in the figure is the EC template).

\subsection{Equilibrium and boundary conditions}\label{sec:equil}

\begin{figure}
    \centering
    \includegraphics[width=.4\textwidth]{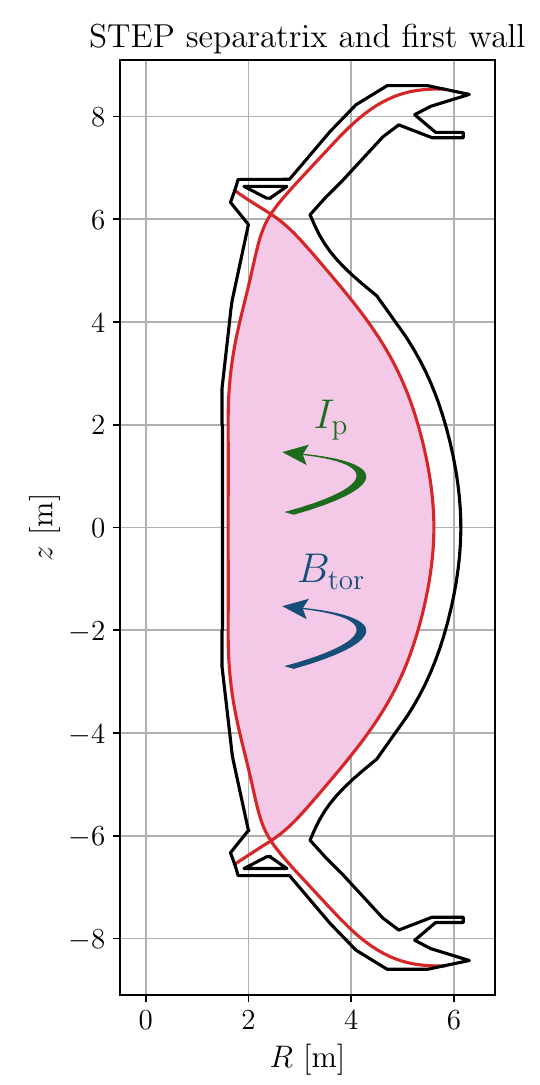}
    \caption{Separatrix geometry assumed for the JETTO simulations, generated with the free equilibrium solver Fiesta~\cite{fiesta}. Arrows show the orientation of the toroidal magnetic field and the plasma current (both are counterclockwise when viewed from above).}
    \label{fig:separatrix}
\end{figure}

A fixed separatrix geometry is assumed in JETTO (see Fig.~\ref{fig:separatrix}), which has been calculated by the free boundary equilibrium solver Fiesta~\cite{fiesta} from a given coil-set geometry. Since JETTO is a core plasma model, particle and power fluxes cannot be evaluated beyond the separatrix. SOLPS-ITER~\cite{solps} is used to predict the SOL plasma and fluxes on plasma facing components and the divertors. Densities and temperatures are given as boundary conditions at the separatrix, with the values $T_\mathrm{e,sep} = 160$\,eV, $T_\mathrm{i,sep} = 400$\,eV, $n_\mathrm{D,sep} = 1.5\times 10^{19}$\,m$^{-3}$, $n_\mathrm{T,sep} = 1.5\times 10^{19}$\,m$^{-3}$. These values have been suggested from iterations between JETTO and SOLPS-ITER. In later stages of the STEP modelling, JETTO and EDGE2D/EIRENE~\cite{edge2d,eirene} will be integrated at runtime via the JINTRAC~\cite{jintrac} integrated modelling framework, which will give more accurate estimates of the separatrix densities and temperatures via direct interaction between the core and the SOL plasmas. There are differences in how the impurity densities and their boundary conditions are handled between the scans and the candidate flat-top points, which is discussed in Sec.~\ref{sec:imprad}. Electron density simply follows from quasi-neutrality assumption. The magnetic equilibrium inside the separatrix is calculated self-consistently in JETTO using the ESCO~\cite{jetto} 2D Grad-Shafranov solver. 

\subsection{Transport and confinement}\label{sec:transp}

The expected transport and overall confinement of the plasma in STEP relevant regimes are presently unknown. Available empirical scalings lack validity in these regimes, and extrapolation of them arrives at different conclusions regarding the confinement and its dependencies with respect to key plasma parameters. Gyrokinetic modelling has shed some light on the kinds of turbulence that are expected to dominate~\cite{roach,kennedy,giacomin}, but more extensive nonlinear modelling is required to say anything about the overall confinement. With JETTO run in assumption integration mode, the density and temperature regimes are essentially defined as inputs to the code, and fuelling rates and anomalous transport are rescaled in feedback loops until the target regimes have been reached. This way, a broad range of plasma regimes can be assessed in terms of fusion power performance and engineering limitations. The confinement assumption can then be further analysed using some of the available empirical scaling laws. The fuelling rate is adapted to reach a target Greenwald density fraction $f_\mathrm{GW} = \langle n_\mathrm{e}\rangle/n_\mathrm{GW}$ (more details in Sec.~\ref{sec:fuelling}), whereas the anomalous heat and particle diffusivities are rescaled to reach a target normalised beta $\beta_\mathrm{N}$.

Heat and particle transport both have a neoclassical component computed by NCLASS~\cite{nclass}, which has not been rescaled. To get a sensible shape of the anomalous heat diffusivity profiles, the Bohm/gyro-Bohm~\cite{bgb} semi-empirical transport model has been used. The relative electron to ion heat diffusivities are tuned to have dominant electron heat transport, motivated by recent ST experiments~\cite{kaye} and gyrokinetic simulations of relevant regimes~\cite{kennedy,giacomin}. The anomalous heat diffusivities are evaluated according to
\begin{eqnarray}
    \chi_\mathrm{e,anom} &= \alpha_\mathrm{BgB}(0.01\chi_\mathrm{e,B}+50\chi_\mathrm{e,gB}),\\
    \chi_\mathrm{i,anom} &= \alpha_\mathrm{BgB}(0.001\chi_\mathrm{e,B}+\chi_\mathrm{e,gB}),
\end{eqnarray}
where $\alpha_\mathrm{BgB}$ is the transport rescaling parameter, adapted to reach a given target $\beta_\mathrm{N}$. Particle diffusivities are derived from heat diffusivities according to
\begin{eqnarray}
    D_\mathrm{p} &= \xi(\rho_\mathrm{tor})\frac{\chi_\mathrm{e}\chi_\mathrm{i}}{\chi_\mathrm{e}+\chi_\mathrm{i}},\\\xi(\rho_\mathrm{tor}) &= A_1 + (A_2 - A_1)\rho_\mathrm{tor},
\end{eqnarray}
where $A_1 = 1.0$ and $A_2 = 0.3$. The linear weight function $\xi(\rho_\mathrm{tor})$ and its parameter values $A_1$ and $A_2$ correspond to an ad hoc rescaling of the particle diffusivity that agrees with transport modelling of JET experiments. Because of the low ion heat diffusivity, which is essentially at neoclassical levels, the resulting particle diffusivity is also almost purely neoclassical. There is an additional inward particle pinch term,
\begin{equation}
    v_\mathrm{in} = 0.5\frac{D_\mathrm{p} S_\mathrm{flux}^2}{V}\left(\frac{\mathrm{d}V}{\mathrm{d}\rho_\mathrm{tor}}\right)^{-1}\sqrt{\frac{\Psi_\mathrm{tor,sep}}{\pi B_\mathrm{geo}}},
\end{equation}
which has also been observed to agree with Bohm/gyro-Bohm modelling of JET experiments. Here, $S_\mathrm{flux}$ is the flux surface area, $V(\rho_\mathrm{tor})$ is the volume contained inside $\rho_\mathrm{tor}$, and $\Psi_\mathrm{tor,sep}$ is the toroidal magnetic flux at the separatrix, offset such that $\Psi_\mathrm{tor,ax} = 0$.

\subsection{Fuelling}\label{sec:fuelling}

The target Greenwald density fraction is reached by feedback on the particle source rate, mimicking pellet fuelling with a simple source density that is Gaussian in $\rho_\mathrm{tor} = \sqrt{\Psi_\mathrm{tor}/\Psi_\mathrm{tor,sep}}$ and continuous in time. The Gaussian is centered at $\rho_\mathrm{tor} = 0.8$, with a width $\sigma_\rho = 0.24/\sqrt{2} \approx 0.17$. The depth and width of the pellet deposition has been guided by calculations with HPI2~\cite{hpi2} discrete pellet model using realistic pellet injector location and geometry.

\subsection{Pedestal}\label{sec:pedestal}

Based on Europed~\cite{europed} modelling of STEP relevant scenarios, the electron pressure pedestal height is scaled according to
\begin{equation}\label{eq:ped}
    \fl p_\mathrm{e,ped}[\mathrm{Pa}] = c_p(n_\mathrm{e,ped}[10^{19}\,\mathrm{m}^{-3}])^{0.4}(I_\mathrm{p}[\mathrm{MA}])^{1.44}(B_\mathrm{geo}[\mathrm{T}])^{0.41}(R_\mathrm{geo}[\mathrm{m}])^{-2.1}(\delta[~])^{1.69},
\end{equation}
where $c_p = 2.728\times 10^3$, and $\delta$ is the triangularity of the last closed flux surface. The ion pressure is assumed to follow the same scaling. The pedestal width in normalised poloidal flux follows the scaling
\begin{equation}\label{eq:pwid}
    \Delta_\mathrm{ped} = 0.1\beta_\mathrm{pol,ped}^{1/2},
\end{equation}
where $\beta_\mathrm{pol,ped} = 4\mu_0 p_\mathrm{e,ped}/\langle B_\mathrm{pol}\rangle^2$ is the poloidal beta, evaluated at the pedestal. The width scaling of eq.~(\ref{eq:pwid}) is the same as in~\cite{snyder}, but with a prefactor 0.1 instead of 0.076.

\subsection{Heating and current drive}\label{sec:hcd}

Since the steady-state current is provided entirely by auxiliary heating systems and the bootstrap current, the total plasma current is an output of the simulations. The bootstrap component is predicted with NCLASS~\cite{nclass} neoclassical model. Alpha heating, including split of heating fraction between electrons and ions, is based on an analytic model described in~\cite{mikkelsen}, and the model for slowing-down distribution with the resulting alpha particle pressure is described in~\cite{estrada}.

At the template simulations for the parameter scans, the EC and EBW heating and current density profiles have been chosen to agree with results from beam-tracing codes GRAY~\cite{gray} and GENRAY~\cite{genray}, and kinetic Fokker--Planck solver CQL3D~\cite{cql3d}. None of these codes are used at runtime during the scans; GENRAY and CQL3D have not been integrated with JETTO, and GRAY is too computationally demanding to use for all (more than 1\,400) scan points. Instead, the codes have been used in a narrow parameter space around the template points to understand the trends of heating and current drive with respect to plasma parameters, which have then been translated to simplified assumptions that are used by JETTO in the scans. The spatial dependence of the heating power density is kept fixed in the scans, with an overall scaling factor proportional to the power of the auxiliary system. The dependencies of the current drive with respect to auxiliary power and plasma parameters are handled differently for the EC and EBW systems. 

The flux surface averaged toroidal EC current density in JETTO is evaluated according to
\begin{equation}\label{eq:jec}
    J_\mathrm{EC}[\mathrm{A}/\mathrm{m}^2] = \zeta_\mathrm{CD}P_\mathrm{EC}[\mathrm{MW}]\Sigma[\mathrm{m}^{-2}]\frac{T_\mathrm{e}[\mathrm{eV}]}{n_\mathrm{e}[10^{19}\,\mathrm{m}^{-3}]},
\end{equation}
where $\zeta_\mathrm{CD}$ is a global parameter quantifying the current drive efficiency, and $\Sigma$ is a factor defined from the EC power density $p_\mathrm{EC}$ and the cross-sectional flux surface area $S$ as
\begin{equation}
    \Sigma(\rho_\mathrm{tor}) = \frac{\rho_\mathrm{tor}p_\mathrm{EC}}{\mathrm{d}S/\mathrm{d}\rho_\mathrm{tor}}\left(\int_0^1\mathrm{d}\rho_\mathrm{tor}\,\rho_\mathrm{tor}p_\mathrm{EC}(\rho_\mathrm{tor})\right)^{-1}.
\end{equation}
The variation in current drive efficiency due to changes in the velocity space interaction of collisions is captured in the $T_\mathrm{e}/n_\mathrm{e}$ dependence of eq.~(\ref{eq:jec}). For the EC template, $\zeta_\mathrm{CD}$ is set to 18.2\footnote{Note that $\zeta_\mathrm{CD}$ has the unit $10^{13}\,\mathrm{A}\,\mathrm{eV}^{-1}\,\mathrm{W}^{-1}\,\mathrm{m}^{-3} \approx 6.242\times 10^{31}\,\mathrm{s}^5\,\mathrm{m}^{-7}\,\mathrm{kg}^{-2}\,\mathrm{A}$, and $\tilde{c}_0$ has the unit $10^{-13}\,\mathrm{A}^{-1}\,\mathrm{eV}\,\mathrm{W}\,\mathrm{m}^2 \approx 1.602\times 10^{-32}\,\mathrm{s}^{-5}\,\mathrm{m}^6\,\mathrm{kg}^2\,\mathrm{A}^{-1}$. When presenting numerical values of $\zeta_\mathrm{CD}$ and $\tilde{c}_0$, these units have been omitted for simplicity.}, whereas the EC+EBW template uses $\zeta_\mathrm{CD} = 22.7$ for the EC current drive. These values are in quantitative agreement with results from GRAY~\cite{gray} current drive modelling. Comparing this to the more conventional definition of the current drive efficiency $\tilde{\zeta}_\mathrm{CD}$ used in the literature~\cite{eccd}, it can be shown that
\begin{equation}\label{eq:zcon}
    \tilde{\zeta}_\mathrm{CD} = \tilde{c}_0R_\mathrm{geo}[\mathrm{m}]\left[\frac{n_\mathrm{e}}{T_\mathrm{e}}\left(\int_0^1\mathrm{d}\rho_\mathrm{tor}\,\rho_\mathrm{tor}p_\mathrm{EC}\right)^{-1}\int_0^1\mathrm{d}\rho_\mathrm{tor}\frac{\rho_\mathrm{tor}p_\mathrm{EC}T_\mathrm{e}}{n_\mathrm{e}}\right]\zeta_\mathrm{CD},
\end{equation}
where $\tilde{c}_0 \approx 3.27\times 10^{-3}$\footnotemark[\value{footnote}]. Note that $\tilde{\zeta}_\mathrm{CD}$ is a local quantity (a function of $\rho_\mathrm{tor}$), taking values corresponding to a scalar times $n_\mathrm{e}/T_\mathrm{e}$. Its value has the most relevance in the parts of the plasma where the EC current is driven, which is also where the power density $p_\mathrm{EC}(\rho_\mathrm{tor})$ becomes large. Assuming for simplicity that the power density can be described by a single localised peak in $\rho_\mathrm{tor}$, the expression within the square brackets of eq.~(\ref{eq:zcon}) becomes close to 1 in the volume of the peak (it becomes exactly 1 when $p_\mathrm{EC}$ is described by a single Dirac delta function in $\rho_\mathrm{tor}$). A global current drive efficiency parameter, here referred to as $\tilde{\zeta}^*_\mathrm{CD}$, can then be defined with a close relationship to the JETTO parameter $\zeta_\mathrm{CD}$ according to
\begin{equation}\label{eq:zco2}
    \tilde{\zeta}_\mathrm{CD}^* = \tilde{c}_0R_\mathrm{geo}\zeta_\mathrm{CD}.
\end{equation}
This parameter takes the values 0.215 and 0.268 for the two templates, respectively. 

Each parameter scan in Section~\ref{sec:scans} assumes the same value of $\zeta_\mathrm{CD}$ (and $\tilde{\zeta}_\mathrm{CD}^*$ by extension) as in the corresponding scan template. This assumption is valid given that the EC wave is absorbed in the same way for all scan points. A violation of this assumption would for instance be if the wave passes through a different set of harmonic resonances and cut-off layers before being absorbed at the target resonance. Scan points where the on-axis density exceeds the relevant cut-off density are flagged as being outside operational regimes (see details in Section~\ref{sec:ecebw}). 

The EBW current drive density profile used in the EC+EBW template have been taken from results with GENRAY/CQL3D. Even though the EBW current drive density does not strictly follow eq.~(\ref{eq:jec}), the area integrated version of the equation can be used in combination with eq.~(\ref{eq:zco2}) to derive an approximation of the normalised current drive efficiency for the EBW current drive to give an idea of how the efficiency compares to ECCD, which yields $\zeta_\mathrm{CD} \approx 72$, or $\tilde{\zeta}_\mathrm{CD} \approx 0.85$. However, $\zeta_\mathrm{CD}$ cannot be assumed to be constant throughout the scan space, since the underlying physics is very different from EC heating and current drive. Studies of EBW current drive efficiency using GENRAY/CQL3D in a local density and temperature domain ($\sim 10$\,\% variation) around the EC+EBW template point at a fixed EBW power resulted in a nearly constant EBW current. The scans that were based on the EC+EBW template (Sections~\ref{sec:scan2} and \ref{sec:scan3}) assume that $I_\mathrm{EBW}/P_\mathrm{EBW}$ is fixed at the template value of 32.89\,kA/MW. 

\subsection{Impurities and radiation}\label{sec:imprad}

The main impurities expected for STEP core plasmas are helium, xenon and argon. Helium results as ash from the D/T fusion reactions. Xenon is introduced via doped fuelling pellets to increase core radiation to the levels required for detachment access. Argon is likewise injected to assist detachment access, which is done by gas puffing in the divertor region (see details in Sec.~\ref{sec:ploss}). The candidate flat-top points of Sec.~\ref{sec:scen} model impurity densities and radiation self-consistently. Impurity transport is handled the same way as the main ions (see Sec.~\ref{sec:transp}), with the neoclassical transport expected to dominate. Atomic physics and the resulting impurity radiation is modelled by SANCO~\cite{sanco}.

Simplifying assumptions are made for the impurity densities and radiation during the scans, since self-consistent treatment of the impurities is computationally demanding to a degree that is not viable for all 1\,400+ scan points. Impurity radiation is prescribed via the radiation fraction $f_\mathrm{rad} = P_\mathrm{rad}/(P_\mathrm{aux} + P_\alpha + P_\mathrm{ohm})$, which is set to 70\,\% in all scan points unless stated otherwise. The radiation density is constant in space ($q_\mathrm{rad} = P_\mathrm{rad}/V$). Xenon and helium impurities are included in the core to account for dilution of the plasma. The impurity densities are set by a fixed $Z_\mathrm{eff} = 2.8$ and a fixed helium to xenon ratio $n_\mathrm{He}/n_\mathrm{Xe} = 152$, which matches the expected dilution from self-consistent impurity modelling in order of magnitude.

\section{Parameter scans}\label{sec:scans}
The chosen scan parameters, which are all defined as scalar inputs to JETTO, are
\begin{enumerate}
    \item Greenwald density fraction $f_\mathrm{GW} = \langle n_\mathrm{e}\rangle_\mathrm{line}/n_\mathrm{GW}$,
    \item Normalised beta $\beta_\mathrm{N}$,
    \item ECCD current drive efficiency $\zeta_\mathrm{CD}$
    \item Auxiliary power ($P_\mathrm{EC}$ and $P_\mathrm{EBW}$),
    \item Radiation fraction $f_\mathrm{rad} = P_\mathrm{rad}/P_\mathrm{tot}$.
\end{enumerate}
Table~\ref{tab:scans} summarises the scan parameters and their ranges in each of the scans. All scan parameters span 21 equally spaced grid points, except for the scan in $\zeta_\mathrm{CD}$, which had 6 grid points. This totals to $21\times 21\times 3 + 21\times 6 = 1\,449$ JETTO runs.
\begin{table}[ht]
    \centering
    \caption{Scan parameters and ranges by section where the scan is presented. Numbers in bold are the scan variables of the particular scan.\vspace{2mm}}
    \label{tab:scans}
    \begin{tabular}{cccccccc}\hline
        Section & Template & $f_\mathrm{GW}$ [\%] & $\beta_\mathrm{N}$ [ ] & $\zeta_\mathrm{CD}$ & $P_\mathrm{EC}$ [MW] & $P_\mathrm{EBW}$ [MW] & $f_\mathrm{rad}$ [\%] \\\hline
       \textbf{\ref{sec:scan1}} & EC & \textbf{50 -- 120} & \textbf{4.2 -- 5.5} & 18.2 & 150 & 0 & 70 \\
       \textbf{\ref{sec:eccdnorm}} & EC & \textbf{50 -- 120} & 4.4 & \textbf{16 -- 27} & 150 & 0 & 70 \\
       \textbf{\ref{sec:scan2}} & EC+EBW & 100 & 3.93 & 22.7 & \textbf{2.5 -- 90} & \textbf{42 -- 140} & 70 \\
       \textbf{\ref{sec:scan3}} & EC+EBW & 100 & 3.93 & 22.7 & 9.0 & \textbf{42 -- 140} & \textbf{20 -- 70} \\\hline
    \end{tabular}
\end{table}

The scan parameters $f_\mathrm{GW}$ and $\beta_\mathrm{N}$ can be set to access a wide range of densities and temperatures, which are key parameters in the fusion power performance. As explained in Sections~\ref{sec:transp} and \ref{sec:fuelling}, respectively, both parameters are defined as targets in feedback loops, with pellet fuelling rate and artificial rescaling of the anomalous diffusivity as control parameters, respectively. Effectively, varying $\beta_\mathrm{N}$ while keeping $f_\mathrm{GW}$ constant approximately varies the temperature while keeping the density constant, assuming small variation in the plasma current. This can be understood from the fact that $\beta_\mathrm{N} \propto \langle p \rangle/I_\mathrm{p} \sim \langle n\rangle \langle T\rangle/I_\mathrm{p} \propto \langle T\rangle f_\mathrm{GW}$ ($n$/$T$ can be either electron or ion density/temperature in this context). Varying $f_\mathrm{GW}$ while keeping $\beta_\mathrm{N}$ constant at approximately constant current varies the density and temperature inversely. This way of accessing different temperature regimes using different transport assumptions rather than varying the auxiliary heating makes the confinement factor $H_{98}^*$ a critical parameter for assessing the viability of the scenario. Considering that the O-X-B coupling scheme is sensitive to the density scale length $L_n = n/|\nabla n|$~\cite{freethy}, the EBW heating and current drive assumptions cannot be straightforwardly extrapolated from the EC+EBW template simulation in density and temperature space. For sufficient confidence in the EBW assumptions, the scan spaces should be restricted to the density and temperature ranges that have already been analysed with GENRAY/CQL3D ($\sim 10$\,\% variation). For this reason, a 2D scan in $f_\mathrm{GW}$ and $\beta_\mathrm{N}$ has been done for the EC template only, as presented in Section~\ref{sec:scan1}.

The assessment of the confinement assumption is studied in more detail with the scan of Section~\ref{sec:eccdnorm}. Confinement scalings other than IPB98(y,2) are evaluated in a broad parameter space to compare the conclusions that they arrive at with regards to minimisation of the confinement factor. The scan spans the same space in Greenwald fraction $f_\mathrm{GW}$ as the one presented in Section~\ref{sec:scan1}, with $\beta_\mathrm{N}$ fixed at 4.4, using the EC template. In addition, the JETTO definition of the EC current drive efficiency $\zeta_\mathrm{CD}$ is scanned in order to access different current drive regimes. 

Exploring the combined parameter space of $P_\mathrm{EC}$ and $P_\mathrm{EBW}$ is of interest because it efficiently varies the current and pressure peakings, with the EC and EBW absorption being on-axis and off-axis, respectively. Performing this scan while keeping the target $f_\mathrm{GW}$ and $\beta_\mathrm{N}$ fixed ensures little variation in average density and temperature, assuming a high bootstrap current fraction that keeps the total plasma current relatively constant, despite the auxiliary current drive varying significantly. At high auxiliary power, the anomalous heat diffusivity can be scaled up while maintaining the set target $\beta_\mathrm{N}$. If the corresponding reduction in energy confinement time $\tau^*_\mathrm{E}$ is larger than the reduction prescribed by the ITER scaling, $\tau^*_\mathrm{98} \propto (P_\mathrm{net}^*)^{-0.69}$, the resulting confinement factor also decreases. A combined scan of $P_\mathrm{EC}$ and $P_\mathrm{EBW}$ for the EC+EBW template is presented in Section~\ref{sec:scan2}.

Similarly to an increase of the auxiliary power, a reduction of the impurity radiation can potentially reduce the confinement factor. On the other hand, $P_\mathrm{sep}$ scales negatively with impurity radiation, meaning that low radiation fraction also increases $P_\mathrm{sep}/R_\mathrm{geo}$. In the scan presented in Section~\ref{sec:scan3}, the combined $f_\mathrm{rad} = P_\mathrm{rad}/P_\mathrm{tot}$ and $P_\mathrm{EBW}$ space is explored, using the EC+EBW template.
\subsection{Greenwald fraction and normalised beta}\label{sec:scan1}

\begin{figure}\centering
\includegraphics[width=\textwidth]{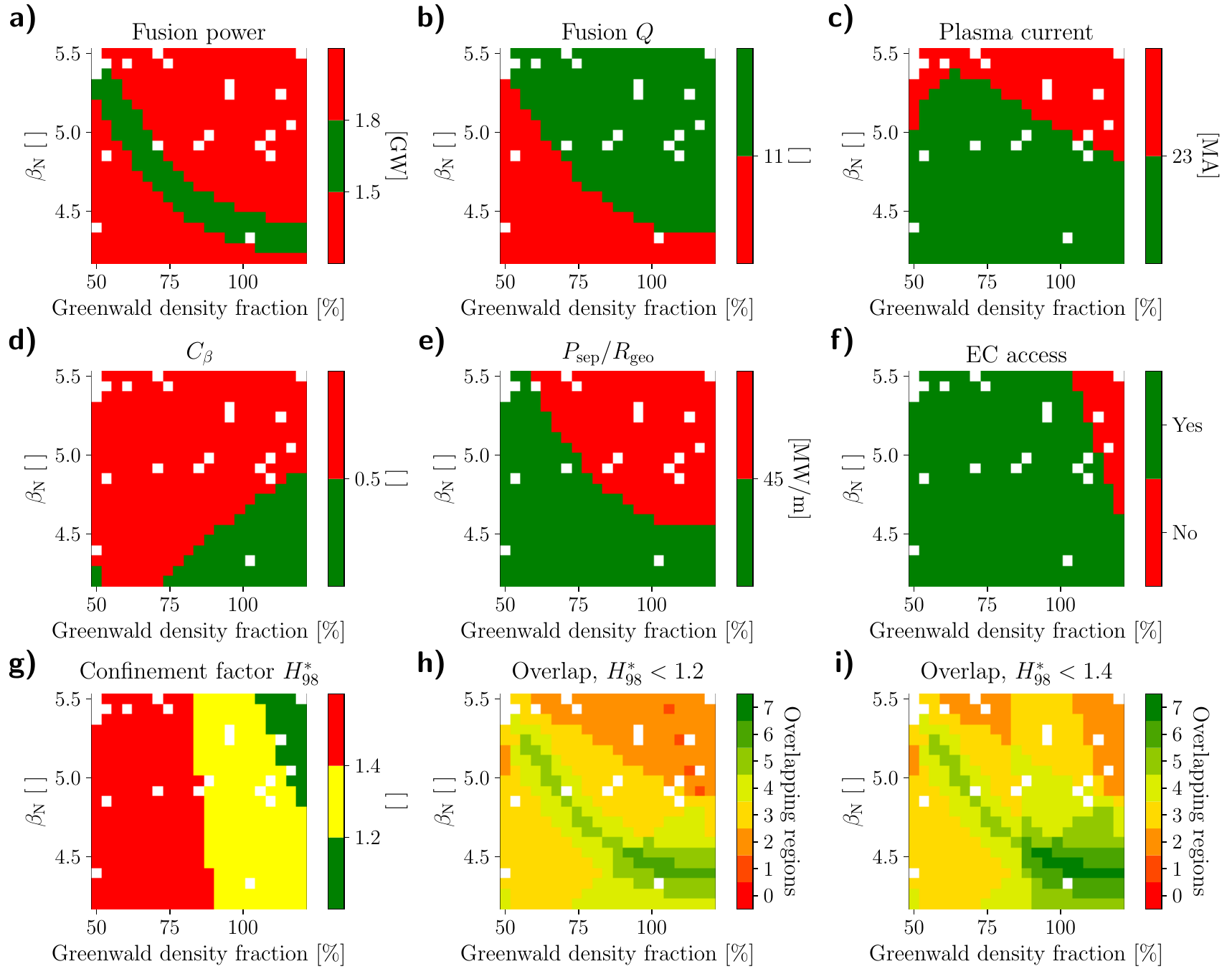}
\caption{2D scan in Greenwald density fraction and normalised beta, based on the EC template simulation. Figures \ref{fig:scan1}.a -- \ref{fig:scan1}.g compares the outputs of all simulations against the listed operational limits in Section~\ref{sec:limsum}, where green means that it operates within limits, and red means that the limit is exceeded. Fig.~\ref{fig:scan1}.g also shows in yellow the parameter space where the optimistic confinement limit is satisfied ($H_{98}^* \leq 1.4$), but not the conservative limit ($H_{98}^* \leq 1.2$). Figures \ref{fig:scan1}.h and \ref{fig:scan1}.i count the total number of operational limits satisfied, where Fig.~\ref{fig:scan1}.h compares against the conservative confinement limit, and Fig.~\ref{fig:scan1}.i compares against the optimistic limit. White dots correspond to gaps in the scan results due to simulations that did not converge for numerical reasons.}
\label{fig:scan1}
\end{figure}

\begin{figure}\centering
\includegraphics[width=\textwidth]{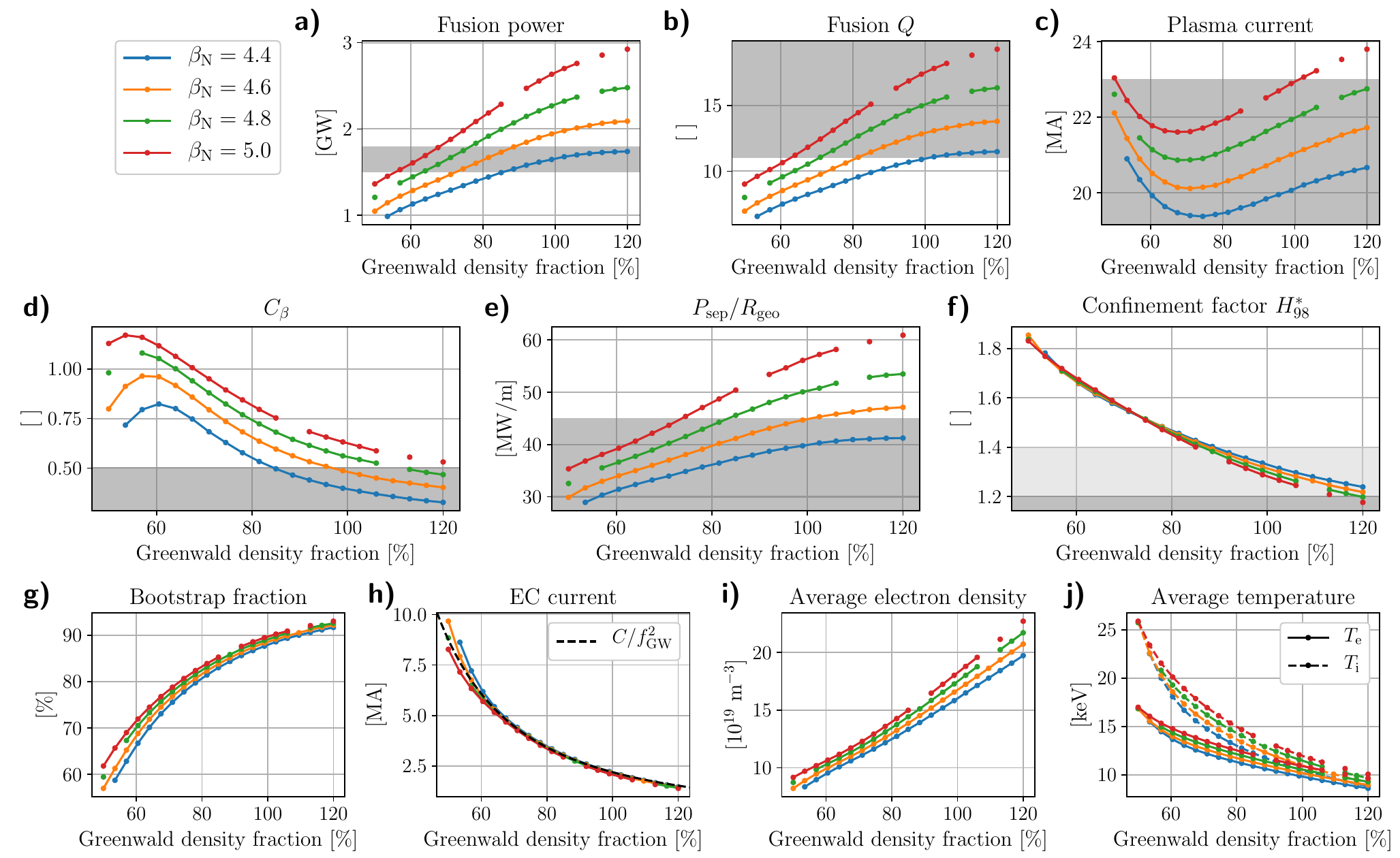}
\caption{Simulation data from the scan in Greenwald density fraction $f_\mathrm{GW}$ and normalised beta $\beta_\mathrm{N}$, sliced at selected constant $\beta_\mathrm{N}$ values. The figures~\ref{fig:s11d}.a -- \ref{fig:s11d}.f present the output parameters that are compared against the set operational limits, with the operational domain of each output highlighted in grey. Figure~\ref{fig:s11d}.f categorises two different operational limits for $H_{98}^*$: The conservative limit $H_{98}^* < 1.2$ shown in dark grey, and the optimistic limit $1.2 \leq H_{98}^* < 1.4$ shown in light grey. Additional simulation output is shown in Figs.~\ref{fig:s11d}.g -- \ref{fig:s11d}.j.}
\label{fig:s11d}
\end{figure}

A full set of 2D plots of the estimated operational domains in the $f_\mathrm{GW}$,$\beta_\mathrm{N}$-space is shown in Fig.~\ref{fig:scan1}, where simulation outputs are compared against the individual limits listed in Section~\ref{sec:limsum} (Figs.~\ref{fig:scan1}a -- \ref{fig:scan1}.g), as well as presenting the overlap of domains by counting number of satisfied operational criteria in each scan point (Figs.~\ref{fig:scan1}.h and \ref{fig:scan1}.i). The two different overlap figures compare $H_{98}^*$ against the conservative and the optimistic limit, respectively. Figure~\ref{fig:s11d} shows the output from the same 2D scan sliced in a set of 1D plots at constant $\beta_\mathrm{N}$ (outputs are linearly interpolated in $\beta_\mathrm{N}$ where selected values do not fall exactly on a scan grid point).

The first observation to make is that the fusion power range $1.5\,\textrm{GW} < P_\mathrm{fus} < 1.8\,\textrm{GW}$ (see Figs.~\ref{fig:scan1}.a and \ref{fig:s11d}.a) covers a narrow domain in $\beta_\mathrm{N}$, with the domain shifted towards higher $\beta_\mathrm{N}$ at low $f_\mathrm{GW}$. The temperature increases to some extent with increasing $\beta_\mathrm{N}$, but the increase is more substantial towards lower $f_\mathrm{GW}$ in the scan domain, with average ion temperature exceeding 25\,keV at $f_\mathrm{GW} = 50$\,\% (see Fig.~\ref{fig:s11d}.j). However, at decreasing $f_\mathrm{GW}$ and constant $\beta_\mathrm{N}$, the increasing ion temperature is insufficient to balance the decreasing density for maintaining the total fusion power, which is why higher $\beta_\mathrm{N}$ is required at low $f_\mathrm{GW}$.

The criterion $Q_\mathrm{sci} > 11$ is only satisfied in a subdomain of the fusion power domain. The narrow band where both $Q_\mathrm{sci} > 11$ and $P_\mathrm{fus} < 1.8$\,GW can be identified from the green hues in the two overlap plots (Figs.~\ref{fig:scan1}.h and \ref{fig:scan1}.i). The criterion $P_\mathrm{sep}/R_\mathrm{geo} < 45$\,MW/m has a limit that follows a similar shape to the $P_\mathrm{fus}$ and $Q_\mathrm{sci}$ limits in $f_\mathrm{GW}$,$\beta_\mathrm{N}$-space. The whole $P_\mathrm{fus}$ domain is contained in the $P_\mathrm{sep}/R_\mathrm{geo}$ domain, meaning that $P_\mathrm{sep}/R_\mathrm{geo} < 45$\,MW/m is redundant and not further restricting the operational space. However, the chance of detachment access generally increases with decreasing $P_\mathrm{sep}$. Figure~\ref{fig:s11d}.e shows that $P_\mathrm{sep}/R_\mathrm{geo} \geq 40$\,MW/m in the $Q_\mathrm{sci} > 11$ domain. Such large $P_\mathrm{sep}/R_\mathrm{geo}$ might still pose a challenge for divertor design and detachment access.

An interesting feature is the non-monotonic dependency of the plasma current with respect to the Greenwald density fraction $f_\mathrm{GW}$, which is clearly seen in Fig.~\ref{fig:s11d}.c. This is because of opposing dependencies of the bootstrap current and the auxiliary current with respect to $f_\mathrm{GW}$ at constant $\beta_\mathrm{N}$. The EC current scales as $T_\mathrm{e}/n_\mathrm{e}$ according to eq.~(\ref{eq:jec}), meaning that the EC current drive is expected to scale similarly to $f_\mathrm{GW}^{-2}$ ($\langle T_\mathrm{e}\rangle \sim \beta_\mathrm{N}/f_\mathrm{GW}$, $\langle n_\mathrm{e}\rangle \propto I_\mathrm{p}f_\mathrm{GW}$). This scaling is confirmed by the least square fit to the current drive, shown by the black dashed curve in Fig.~\ref{fig:s11d}.h. The increasing bootstrap current with increasing $f_\mathrm{GW}$ can be understood from the fact that the edge pressure gradient increases with increasing average density. It is possible that the fixed boundary density and temperatures contributes to this effect, and that it would become weaker with more self-consistent modelling of both the core and the SOL.

\begin{figure}
    \centering
    \includegraphics[width=\textwidth]{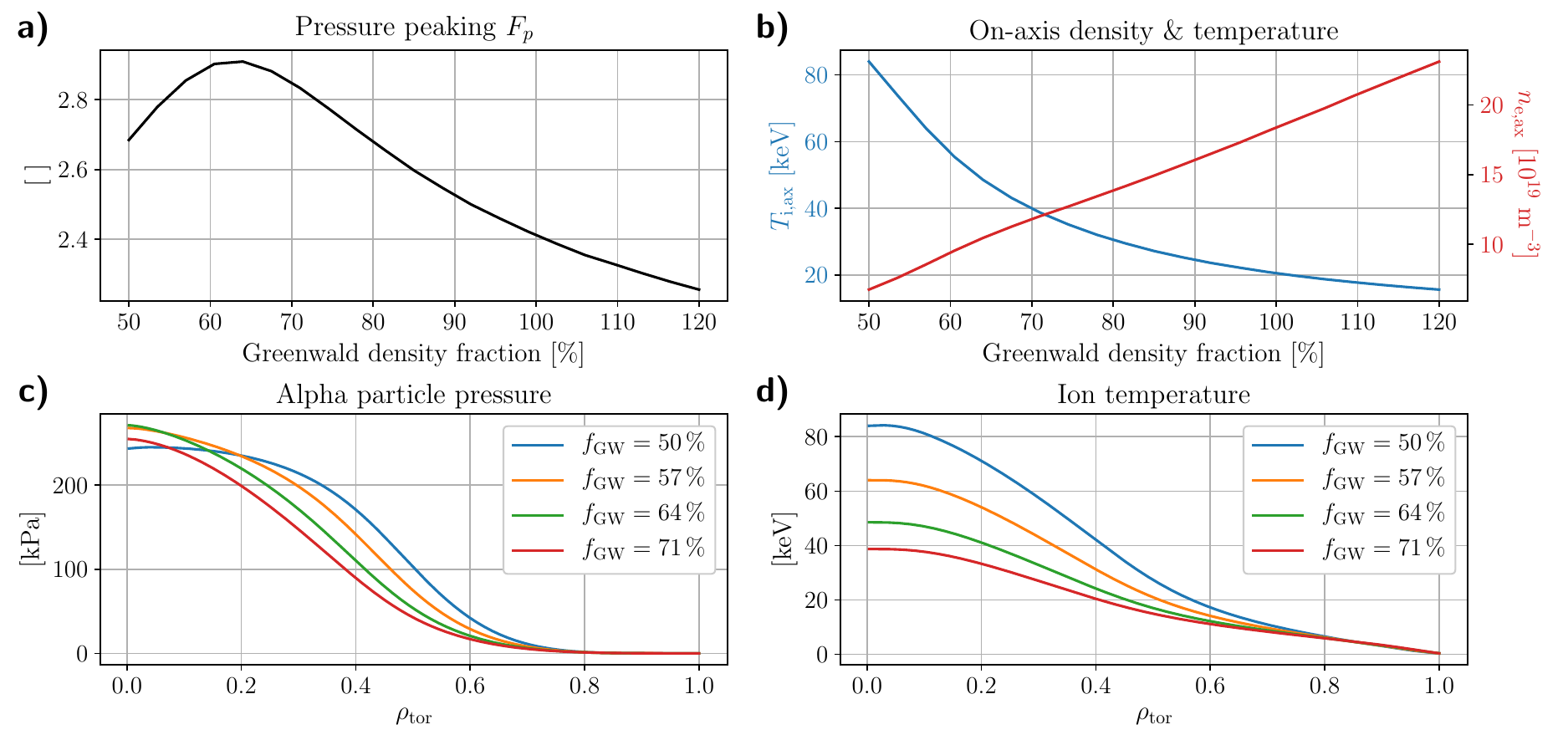}
    \caption{Data from the scan in $f_\mathrm{GW}$ and $\beta_\mathrm{N}$ at $\beta_\mathrm{N} = 4.2$.}
    \label{fig:alpha}
\end{figure}

Similarly to the plasma current, there is a non-monotonicity of $C_\beta$ with respect to $f_\mathrm{GW}$ at constant $\beta_\mathrm{N}$, which indicates that also the pressure peaking is non-monotonic (since $C_\beta$ is a function of $\beta_\mathrm{N}$ and $F_p = p_\mathrm{ax}/\langle p \rangle$ only, see eq.~(\ref{eq:cbeta})). The reduction of pressure peaking at low Greenwald fraction can be understood from changes in the on-axis alpha particle component of the pressure as the ion temperature locally exceeds $\sim$ 50\,keV, with the details demonstrated in Fig.~\ref{fig:alpha}. Above these temperatures, the DT fusion cross section stops increasing significantly with ion temperature, meaning that it is unable to compensate for the decreasing density with decreasing Greenwald fraction to maintain on-axis fusion reaction rates. Unsurprisingly, $C_\beta$ increases with increasing $\beta_\mathrm{N}$. Although the pressure peaking decreases slightly with $\beta_\mathrm{N}$ at constant $f_\mathrm{GW}$, this is not enough to compensate for the explicit $\beta_\mathrm{N}$ dependence in eq.~(\ref{eq:cbeta}).

The conservative $H_{98}^*$ criterion is only satisfied in the high-$f_\mathrm{GW}$, high-$\beta_\mathrm{N}$ corner of the scan space (Fig.~\ref{fig:scan1}.g), which is beyond operational limits in $P_\mathrm{fus}$, $I_\mathrm{p}$ and $P_\mathrm{sep}/R_\mathrm{geo}$. In essence, this indicates that there are no operating points for a non-inductive scenario with sufficient fusion gain and realistic ECCD efficiency assumptions below $H_{98}^* = 1.2$. However, relaxing the confinement requirement to $H_{98}^* < 1.4$, a domain is found where all seven of the assumed operational limits are satisfied, shown with dark green in Fig.~\ref{fig:scan1}.i ($f_\mathrm{GW} \gtrsim 95$\,\%, $\beta_\mathrm{N} \approx 4.4$). As mentioned in Sec.~\ref{sec:econf}, different confinement scalings suggest different density dependencies, where the IPB98(y,2) scaling has a relatively high positive density scaling with $\tau_\mathrm{IPB98(y,2)} \propto \langle n_\mathrm{e}\rangle^{0.41}$. Section~\ref{sec:eccdnorm} compares a slice of the presented scan results ($\beta_\mathrm{N}$ is fixed at 4.4) against a range of confinement scalings. The EC current drive efficiency $\zeta_\mathrm{CD}$ is varied in addition to study its impact on confinement.

The confinement factor is mostly a function of the Greenwald fraction in the scan space, with a non-intuitive dependence with respect to $\beta_\mathrm{N}$ resulting from the non-inductive constraint. To understand how the constraint influences the $\beta_\mathrm{N}$ dependence of the scaling factor, it can first be identified that
\begin{equation}
    \tau_\mathrm{E}^* = \frac{W_\mathrm{th}}{P_\mathrm{net}^*} \sim \frac{\langle p_\mathrm{th}\rangle}{P_\mathrm{net}^*} \sim \frac{I_\mathrm{p} \beta_\mathrm{N,th}}{P_\mathrm{net}^*},~\tau_\mathrm{98}^* \sim I_\mathrm{p}^{0.93}\langle n_\mathrm{e}\rangle^{0.41}(P_\mathrm{net}^*)^{-0.69},
\end{equation}\begin{eqnarray}
    H_{98}^* &= \frac{\tau_\mathrm{E}^*}{\tau_\mathrm{98}^*} \sim I_\mathrm{p}^{0.07}\langle n_\mathrm{e}\rangle^{-0.41}(P_\mathrm{net}^*)^{-0.31}\beta_\mathrm{N,th} \nonumber\\&\sim I_\mathrm{p}^{-0.34}f_\mathrm{GW}^{-0.41}(P_\mathrm{net}^*)^{-0.31}\beta_\mathrm{N,th},
\end{eqnarray}
where $W_\mathrm{th}$, $p_\mathrm{th}$ and $\beta_\mathrm{N,th}$ are the thermal stored energy, pressure and normalised $\beta$, respectively. $P_\mathrm{net}^*$ is the radiation corrected power (see details in Section~\ref{sec:econf}). Fixed field and geometry is assumed, with their dependencies consequently neglected in the derivations. By interpolating the scan data along $f_\mathrm{GW} = 100$\,\% and fitting each parameter $y \in \{I_\mathrm{p},P_\mathrm{net}^*,\beta_\mathrm{N,th}\}$ to a power law dependence in $\beta_\mathrm{N}$ according to $y \approx c_y \beta_\mathrm{N}^{\alpha_y}$ yields the exponents $\alpha_I = 1.03$, $\alpha_P = 2.76$, and $\alpha_\beta = 0.90$. This gives a total $\beta_\mathrm{N}$ dependence in $H_{98}^*$ according to $H_{98}^* \sim \beta_\mathrm{N}^{\alpha_\beta-0.34\alpha_I-0.31\alpha_P} = \beta_\mathrm{N}^{-0.30}$. A corresponding analysis of the dependence of $H_\mathrm{98}^*$ with respect to the Greenwald fraction $f_\mathrm{GW}$ is challenging, since neither of the parameters $I_\mathrm{p}$, $P_\mathrm{net}^*$, and $\beta_\mathrm{N,th}$ follow simple power law dependencies with respect to $f_\mathrm{GW}$ at constant $\beta_\mathrm{N}$. However, making a power law fit to $H_{98}^*(f_\mathrm{GW})$ directly at $\beta_\mathrm{N} = 4.4$ yields an overall exponent $\alpha_f = -0.42 \approx -0.41$, indicating that $I_\mathrm{p}^{-0.34}(P_\mathrm{net}^*)^{-0.31}\beta_\mathrm{N,th}$ is largely a function of $\beta_\mathrm{N}$, independent of $f_\mathrm{GW}$.

The estimated domain of access to the 2$^\mathrm{nd}$ harmonic O-mode for EC heating and current drive is shown in Fig.~\ref{fig:scan1}.f, based on $n_\mathrm{e,ax}$ and $B_\mathrm{ax}$ at each scan point. Since $B_\mathrm{geo}$ is held fixed in the scan, the on-axis magnetic field strength $B_\mathrm{ax}$ only varies with the Shafranov shift. On the other hand, the density covers a wide range in the scan, as indicated by Fig.~\ref{fig:s11d}.i. It is only in the high-density corner of the scan space (high Greenwald fraction and high $\beta_\mathrm{N}$) that the on-axis density exceeds the cut-off density. However, this domain is already outside of the operational limits in fusion power, plasma current and $P_\mathrm{sep}/R_\mathrm{geo}$. Hence, EC coupling access does not restrict the overall operational regime.

\subsection{Greenwald fraction and normalised ECCD efficiency}\label{sec:eccdnorm}

No empirical confinement scaling has a validity domain that includes STEP relevant regimes, since there have never been any experiments of similar scale. The confinement in the scans is mainly assessed by comparing the IPB98(y,2) (radiation-corrected) confinement factor against a set of fixed upper limits. In order for the IPB98(y,2) confinement factor to accurately quantify the degree of over-/under-estimation of the confinement, the confinement time must have similar dependencies with respect to current, density and power around the STEP operating points as those suggested by $\tau_\mathrm{IPB98(y,2)}^* \propto I_\mathrm{p}^{0.93}\langle n_\mathrm{e}\rangle^{0.41}(P_\mathrm{net}^*)^{-0.69}$. The main argument in favour of using IPB98(y,2) for this purpose compared to other empirical scalings is that it is based on H-mode data for a relatively large set of experiments.

As seen in Fig.~\ref{fig:s11d}.f, the IPB98(y,2) scaling indicates that a high Greenwald fraction minimises the confinement factor, which makes operating points in that regime to appear the most realistic from a confinement assumption perspective. In this subsection, a range of alternative confinement factor scalings are evaluated for a 1D slice of the scan of Sec.~\ref{sec:scan1} (at $\beta_\mathrm{N} = 4.4$) in order to test the hypothesis that a high Greenwald fraction is preferable irrespective of the confinement scaling to compare against. In conjunction, the normalised ECCD current drive efficiency $\zeta_\mathrm{CD}$ is scanned to study how different $\zeta_\mathrm{CD}$ assumptions perturbs the domains of minimum confinement. Scanning $\zeta_\mathrm{CD}$ effectively varies the plasma current $I_\mathrm{p}$, which is also a parameter that has a wide range of suggested dependencies according to different empirical scalings.

\begin{figure}
    \centering
    \includegraphics[width=\textwidth]{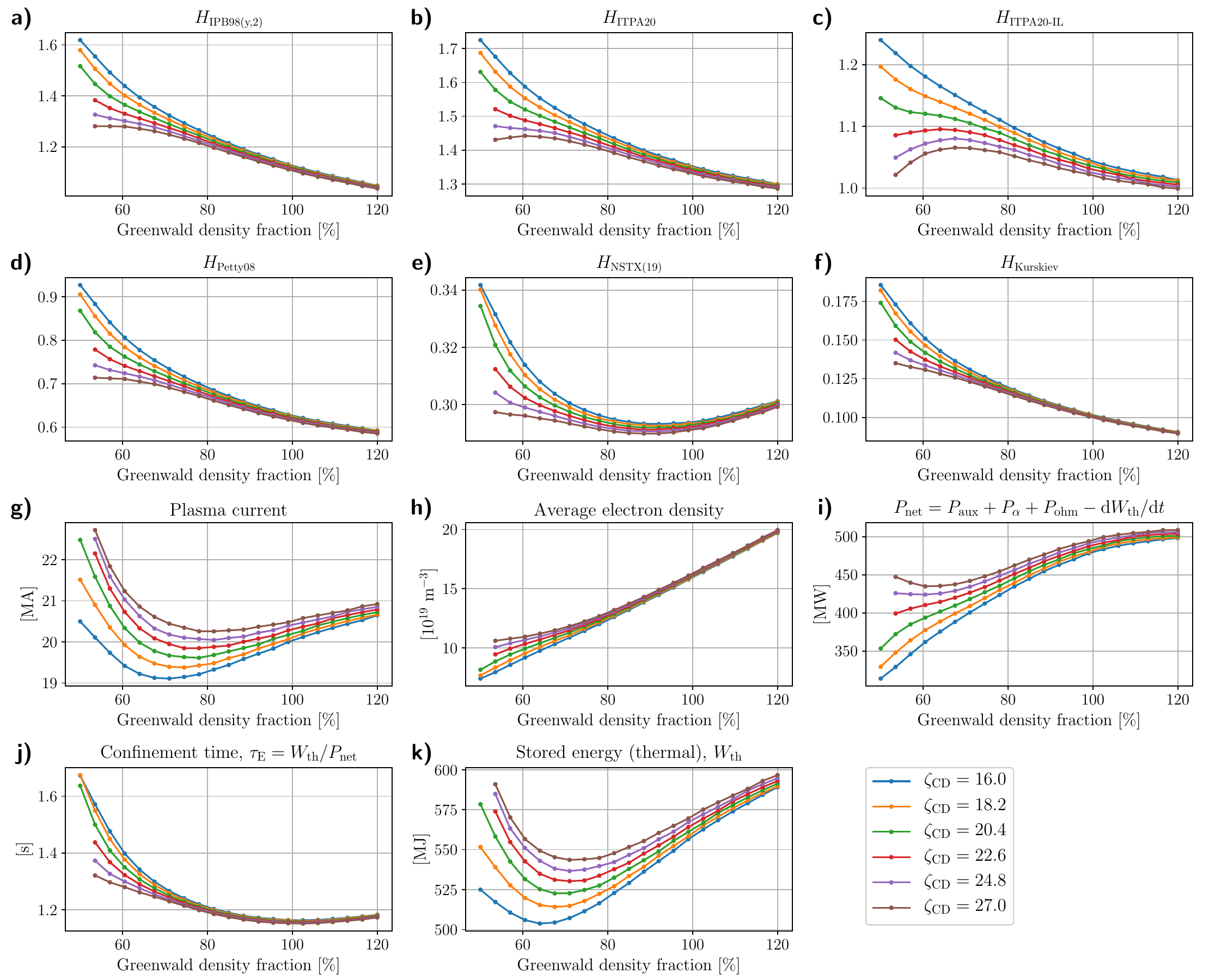}
    \caption{Confinement factors evaluated using a range of different empirical scalings, based on a 2D scan in Greenwald fraction $f_\mathrm{GW}$ and normalised ECCD efficiency $\zeta_\mathrm{CD}$ (see definition in eq.~(\ref{eq:jec})) using the EC only scan template.}
    \label{fig:hfactor}
\end{figure}

\begin{table}[t]
    \centering
    \caption{Exponents for the plasma current, line averaged density and net power according to the confinement scalings presented in Fig.~\ref{fig:hfactor}.}\vspace{2mm}
    \label{tab:hfac}
    \begin{tabular}{ccccccc}\hline
        & \textbf{IPB98(y,2)} & \textbf{ITPA20} & \textbf{ITPA20-IL} & \textbf{Petty08} & \textbf{NSTX(19)} & \textbf{Kurskiev} \\\hline
        $I_\mathrm{p}$ & 0.93 & 0.976 & 1.291 & 0.75 & 0.54 & 0.53 \\
        $\langle n_\mathrm{e}\rangle$ & 0.41 & 0.2442 & 0.1473 & 0.32 & $-$0.05 & 0.65 \\
        $P_\mathrm{net}$ & $-$0.69 & $-$0.6687 & $-$0.6442 & $-$0.47 & $-$0.38 & $-$0.58\\\hline
    \end{tabular}
\end{table}

Figures~\ref{fig:hfactor}.a -- \ref{fig:hfactor}.f evaluate the confinement factor according to the IPB98(y,2)~\cite{iter}, ITPA20~\cite{ITPA20}, ITPA20$-$IL~\cite{ITPA20}, Petty08~\cite{petty}, NSTX(19)~\cite{buxton}, and Kurskiev~\cite{kurskiev} scalings for the scan in $f_\mathrm{GW}$ and $\zeta_\mathrm{CD}$, and Figures~\ref{fig:hfactor}.g -- \ref{fig:hfactor}.k show the input parameters for the scalings that vary across the scan. For consistency, radiation is not taken into account, since the correction for highly core-radiating plasmas has only been derived for the IPB98(y,2) scaling~\cite{lux}. Most scalings predict a monotonically decreasing confinement with increasing Greenwald fraction. One exception is the NSTX(19) scaling, which has a minimum around $f_\mathrm{GW} = 90$\,\%. This is because of the negative density scaling ($\tau_\mathrm{NSTX(19)} \propto \langle n_\mathrm{e}\rangle^{-0.05}$), which together with the negative power scaling causes a net increase of the confinement factor for $f_\mathrm{GW} \gtrsim 90$\,\%. The ITPA20-IL scaling also shows non-monotonicity of the confinement factor with respect to $f_\mathrm{GW}$ at high ECCD efficiency. The relatively strong current dependency of $\tau_\mathrm{ITPA20{\mhyphen}IL} \propto I_\mathrm{p}^{1.291}$ decreases the confinement factor as the ECCD current increases at low $f_\mathrm{GW}$, high $\zeta_\mathrm{CD}$.  As will be seen in Section~\ref{sec:scen}, a high-density flat-top point ($f_\mathrm{GW} \sim 100$\,\%) has been derived as a potential candidate for the EC only scenario, which is partly based on the results indicated by the IPB98(y,2) scaling. A low-density candidate point is suggested as an alternative, which considers results from other confinement scalings (in particular ITPA20-IL). 

\subsection{EBW and EC power}\label{sec:scan2}

\begin{figure}\centering
\includegraphics[width=\textwidth]{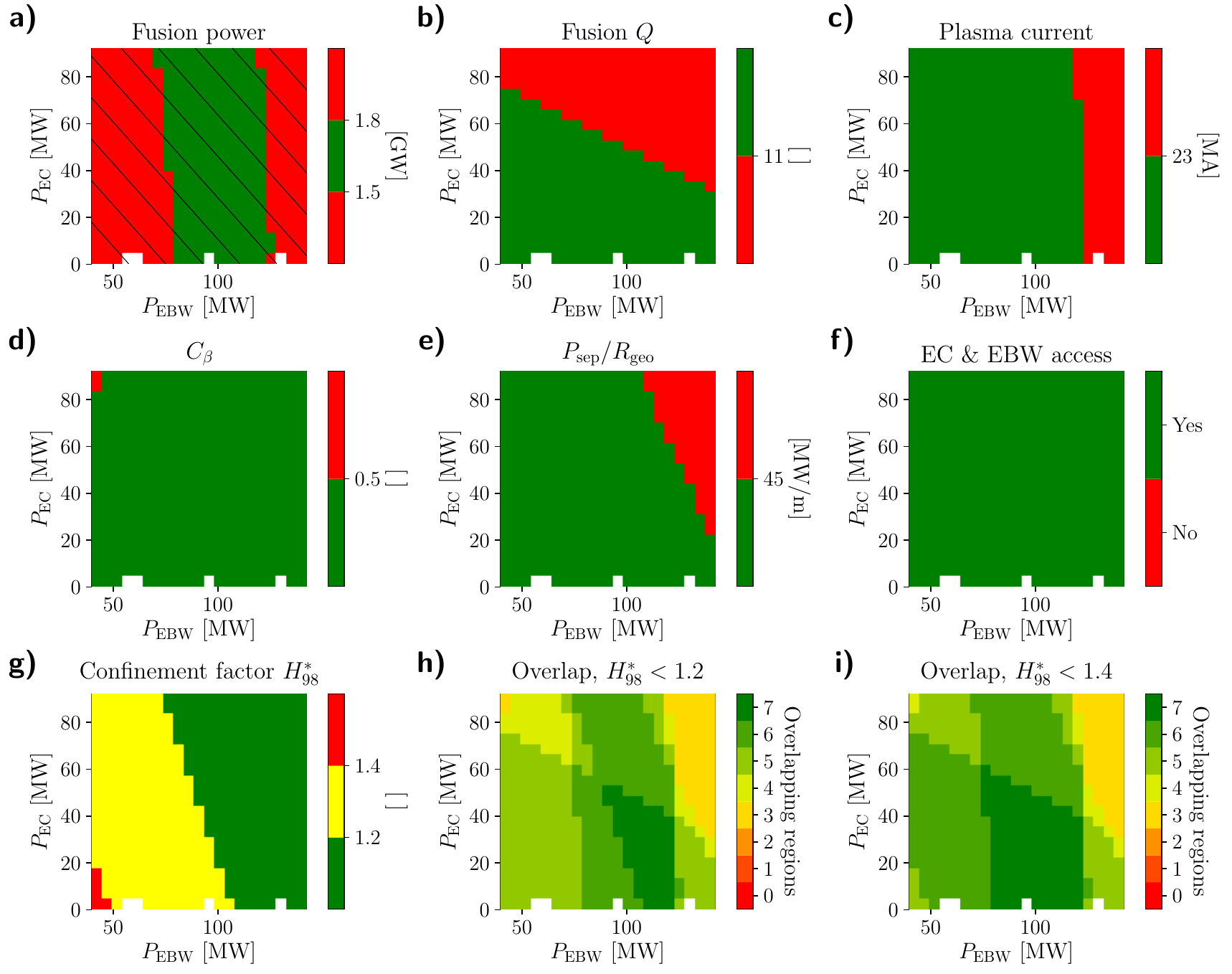}
\caption{2D scan in EC and EBW auxiliary power, based on the EC+EBW template simulation. Figures \ref{fig:scan2}.a -- \ref{fig:scan2}.g compares the outputs of all simulations against the listed operational limits in Section~\ref{sec:limsum}, where green means that it operates within limits, and red means that the limit is exceeded. The black lines in Fig.~\ref{fig:scan2}.a correspond to curves of constant auxiliary power. Fig.~\ref{fig:scan2}.g shows in yellow the parameter space where the optimistic confinement limit is satisfied ($H_{98}^* \leq 1.4$), but not the conservative limit ($H_{98}^* \leq 1.2$). Figures \ref{fig:scan2}.h and \ref{fig:scan2}.i count the total number of operational limits satisfied, where Fig.~\ref{fig:scan2}.h compares against the conservative limit, and Fig.~\ref{fig:scan2}.i compares against the optimistic limit.}
\label{fig:scan2}
\end{figure}

\begin{figure}\centering
\includegraphics[width=\textwidth]{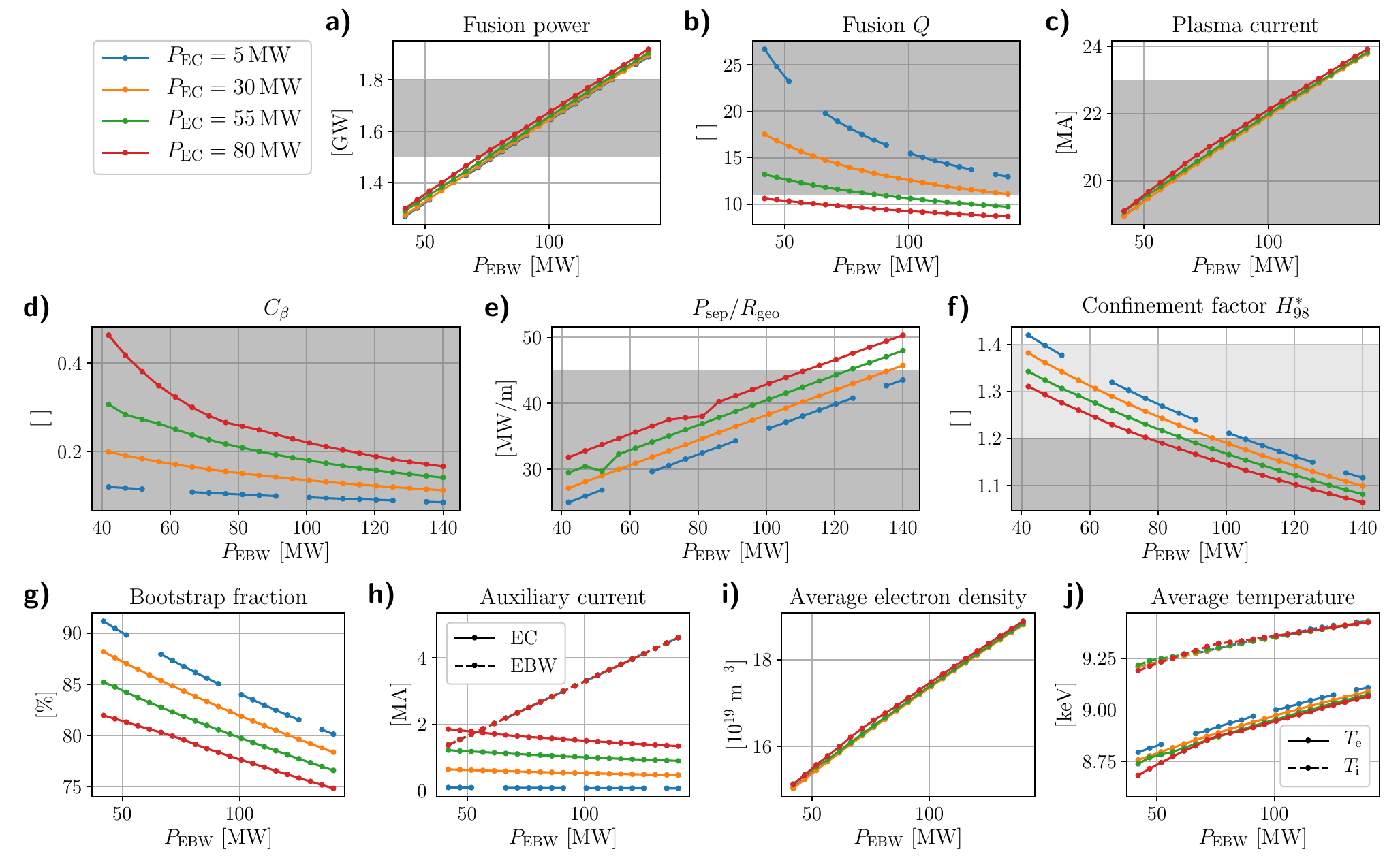}
\caption{Simulation data from the scan in EC and EBW auxiliary power, sliced at selected constant $P_\mathrm{EC}$ values. The figures~\ref{fig:s21d}.a -- \ref{fig:s21d}.f present the output parameters that are compared against the set operational limits, with the operational domain of each output highlighted in grey. Figure~\ref{fig:s21d}.f categorises two different operational limits for $H_{98}^*$: The conservative limit $H_{98}^* < 1.2$ shown in dark grey, and the optimistic limit $1.2 \leq H_{98}^* < 1.4$ shown in light grey. Additional simulation output is shown in Figs.~\ref{fig:s21d}.g -- \ref{fig:s21d}.j.}
\label{fig:s21d}
\end{figure}

The results of the scan in EC and EBW power space are shown in Figs.~\ref{fig:scan2} and \ref{fig:s21d}. The EBW system, which has off-axis heating and current drive, is generally more efficient than EC for driving current. This is supported by the observations in Figs.~\ref{fig:scan2}.c and \ref{fig:s21d}.c. The fusion power also increases significantly more with respect to EBW power than EC power, even though the on-axis EC heating couples directly to the core, where the fusion power density usually peaks. However, the target Greenwald density fraction in the fuelling rate feedback is kept fixed at 100\,\%, meaning that the average density scales with the total plasma current. It can be observed in Fig.~\ref{fig:s21d}.i that the average density indeed increases with the EBW power, which in turn contributes to the increase of fusion power. The fact that the target $\beta_\mathrm{N}$ also remains fixed causes the average temperature to be relatively constant throughout the auxiliary power scan space, as seen in Fig.~\ref{fig:s21d}.j. Unsurprisingly, the fusion $Q$ is the highest in the low auxiliary power corner of the scan.

Comparing the $Q > 11$ domain of Fig.~\ref{fig:scan2}.b against the curves of constant auxiliary power in Fig.~\ref{fig:scan2}.a, EBW heating appears to be preferable over EC heating for high fusion $Q$. $\beta_\mathrm{N}$ is held fixed throughout the scan at a value of 3.93, which is marginally above the estimated no-wall $\beta_\mathrm{N}$ limit of 3.7. $C_\beta$ only exceeds 0.5 for two scan points in the low EBW power, high EC power corner of the scan space, where the pressure peaking is large and the RWM ideal-wall limit is low in $\beta_\mathrm{N}$. The main disadvantages of EBW heating and current drive compared to EC, as observed from the operational domains in Fig.~\ref{fig:scan2}, are the relatively higher $P_\mathrm{sep}/R_\mathrm{geo}$ along the curves of constant auxiliary power (Fig.~\ref{fig:scan2}.e), and the more efficient current drive risking to exceed total current limits (Fig.~\ref{fig:scan2}.c). However, in the auxiliary power scan space, there is a wide region where both $P_\mathrm{sep}/R_\mathrm{geo}$ and $I_\mathrm{p}$ are well within operational limits.

The confinement factor $H_{98}^*$ is the lowest in the regimes of high auxiliary power, which follows from the fact that high heating power can maintain the temperature regime prescribed by the $f_\mathrm{GW}$, $\beta_\mathrm{N}$ combination at low energy confinement time. This reduction of confinement time is larger than the reduction suggested by the $\tau_\mathrm{IPB98(y,2)}$ scaling. Similarly, high auxiliary current drive (primarily with EBW) tends to reduce the confinement time relative to $\tau_\mathrm{IPB98(y,2)} \propto I_\mathrm{p}^{0.93}$. Domains where the conservative confinement assumption is satisfied can be found for auxiliary powers above 113.1\,MW in the scan space, with a preference for high EBW power over high EC power along curves of constant auxiliary power. There is a relatively wide region in the scan space where all seven operational limits are satisfied, including the conservative confinement assumption, as seen in Fig.~\ref{fig:scan2}.h. With the optimistic assumption in Fig.~\ref{fig:scan2}.i, the fully compliant operating space is no longer limited by high $H_{98}^*$, but rather by $P_\mathrm{fus} > 1.5$\,GW (for $P_\mathrm{EBW} \lesssim 80$\,MW). Increasing $\beta_\mathrm{N}$ can potentially bring the operating point above the fusion power lower limit while satisfying the optimistic confinement assumption at low auxiliary power, suggesting the existence of a high fusion $Q$ operating point in the EC+EBW flat-top scenario.

Access to both EC and EBW heating and current drive is estimated in Fig.~\ref{fig:scan2}.f. Red coloured domains would indicate that at least one of the two auxiliary systems lacks access. With neither the density nor the on-axis magnetic field varying significantly from the values at the EC + EBW scan template, which has confirmed EC and EBW access, there is no part of the scan space where the electron gyro-frequency falls outside any of the coupling domains.

\subsection{EBW power and radiation fraction}\label{sec:scan3}

\begin{figure}\centering
\includegraphics[width=\textwidth]{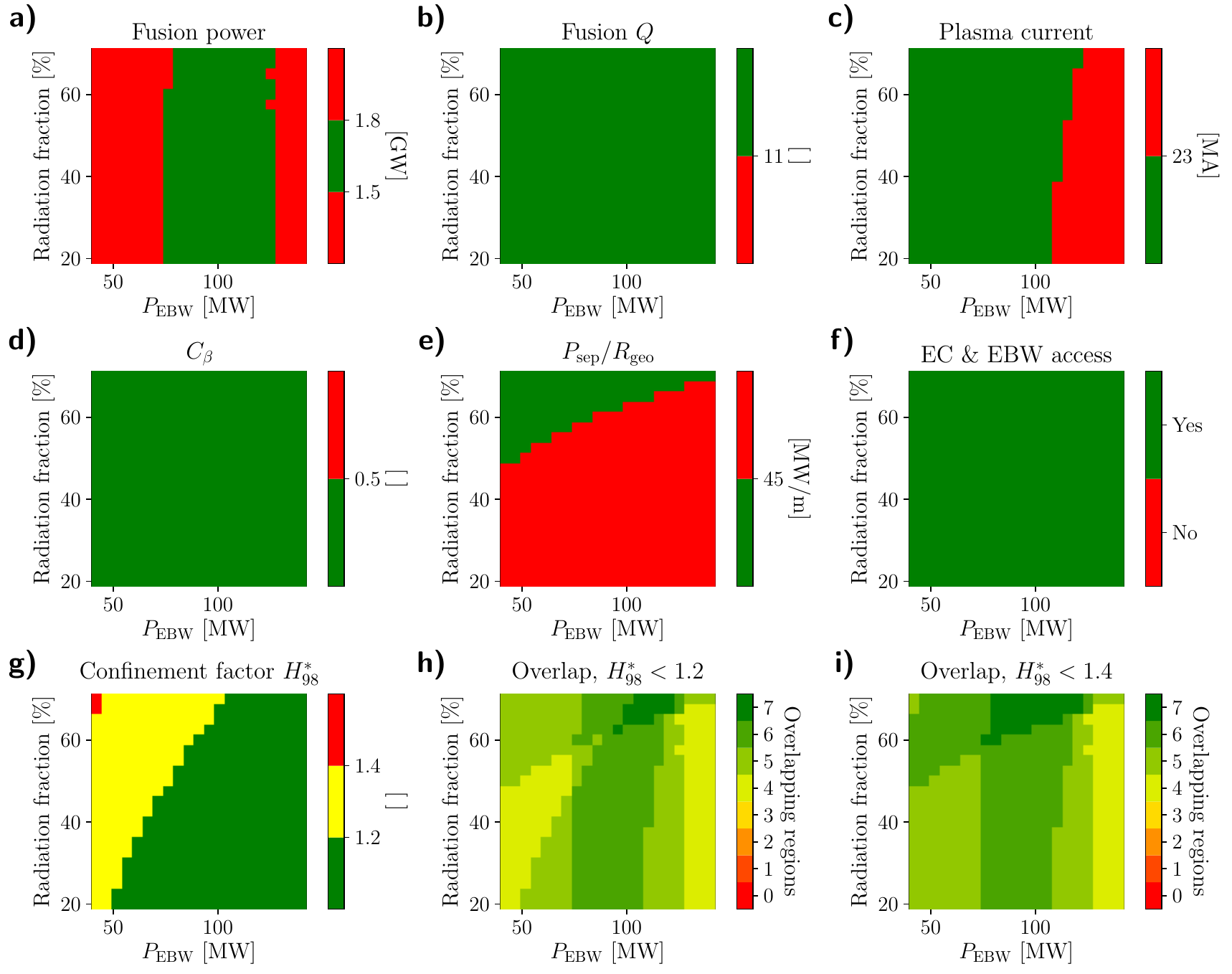}
\caption{2D scan in EBW power and radiation fraction, based on the EC+EBW template simulation. Figures \ref{fig:scan3}.a -- \ref{fig:scan3}.g compares the outputs of all simulations against the listed operational limits in Section~\ref{sec:limsum}, where green means that it operates within limits, and red means that the limit is exceeded. Fig.~\ref{fig:scan3}.g shows in yellow the parameter space where the optimistic confinement limit is satisfied ($H_{98}^* \leq 1.4$), but not the conservative limit ($H_{98}^* \leq 1.2$). Figures \ref{fig:scan3}.h and \ref{fig:scan3}.i count the total number of operational limits satisfied, where Fig.~\ref{fig:scan3}.h compares against the conservative limit, and Fig.~\ref{fig:scan3}.i compares against the optimistic limit.}
\label{fig:scan3}
\end{figure}

\begin{figure}\centering
\includegraphics[width=\textwidth]{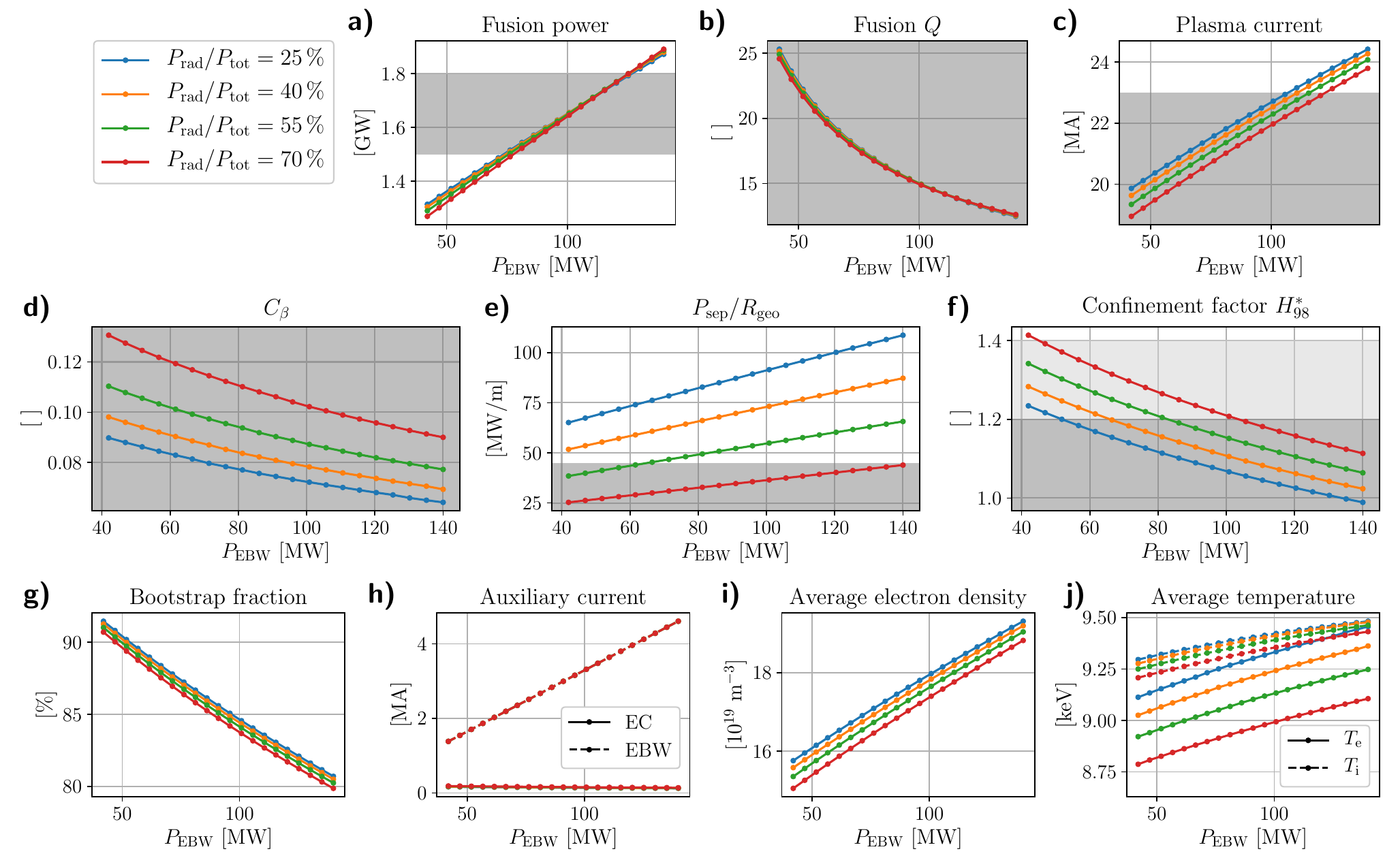}
\caption{Simulation data from the scan in EBW power and radiation fraction, sliced at selected constant $f_\mathrm{rad} = P_\mathrm{rad}/P_\mathrm{tot}$ values. The figures~\ref{fig:s31d}.a -- \ref{fig:s31d}.f present the output parameters that are compared against the set operational limits, with the operational domain of each output highlighted in grey. Figure~\ref{fig:s31d}.f categorises two different operational limits for $H_{98}^*$: The conservative limit $H_{98}^* < 1.2$ shown in dark grey, and the optimistic limit $1.2 \leq H_{98}^* < 1.4$ shown in light grey. Additional simulation output is shown in Figs.~\ref{fig:s31d}.g -- \ref{fig:s31d}.j.}
\label{fig:s31d}
\end{figure}

The results from the scan varying the EBW power and radiation fraction are presented in Figs.~\ref{fig:scan3} and \ref{fig:s31d}. Radiation fraction is not varied self-consistently with core impurity content in this scan. Rather, the core radiation is prescribed, with a flat profile in radiation power density for simplicity (see details of the impurity and radiation assumptions in Sec.~\ref{sec:imprad}). Since the target Greenwald density fraction and the normalised $\beta$ are both kept constant, the average temperatures remain approximately constant in the whole scan space. A reduction of the radiation fraction results in increased transport and lower $H^*_{98}$. Since variation of the radiation fraction also does not influence the plasma current considerably, as seen in Fig.~\ref{fig:s31d}.c., the density also remain relatively unchanged along this scan dimension. For this reason, the fusion power and fusion $Q$ do not depend on the radiation fraction, as demonstrated in Figs.~\ref{fig:s31d}.a and \ref{fig:s31d}.b. 

Similarly to the scan of Section~\ref{sec:scan2}, $\beta_\mathrm{N}$ is fixed at $\beta_\mathrm{N} = 3.93$, which is close to the estimated RWM no-wall limit. With the EC power being fixed at 9.0\,MW, the pressure peaking never gets large enough to significantly reduce the ideal-wall limit and push $C_\beta$ above 0.5. Also, neither the on-axis electron density nor the magnetic field vary significantly in the scan space (see Figs.~\ref{fig:s31d}.i and \ref{fig:s31d}.j), meaning that EC and EBW are both likely to be accessible in the whole scan domain, as indicated by Fig.~\ref{fig:scan3}.f.

The domain where all seven operational limits are satisfied, including the con\-ser\-va\-tive confinement assumption, is primarily limited by $P_\mathrm{sep}/R_\mathrm{geo}$ and $H_{98}^*$, which can be seen in Figs.~\ref{fig:scan3}.e, \ref{fig:scan3}.g and \ref{fig:scan3}.h. With the optimistic confinement assumption, the operational domain is limited in the low end of $P_\mathrm{EBW}$ by $P_\mathrm{fus} > 1.5$\,GW. Due to the $P_\mathrm{sep}/R_\mathrm{geo} < 45$\,MW/m limit, the radiation fraction cannot drop below 60\,\% while still satisfying either the fusion power limitation or the conservative confinement assumption. Increasing $\beta_\mathrm{N}$ for expanding the fusion power limit in the low EBW power domain is un\-like\-ly to expand the operational radiation fraction domain, since increasing $\beta_\mathrm{N}$ also increases $P_\mathrm{sep}/R_\mathrm{geo}$ (see Fig.~\ref{fig:s11d}.e).

\section{Candidate operational points}\label{sec:scen}
The analysis provided by the scans suggests at least four candidate flat-top points, depending on the confinement assumptions: two EC only and two EC+EBW operational points. The EC only scenario suggests $\beta_\mathrm{N} \approx 4.5$ and $f_\mathrm{GW} \approx 100\,\%$ with $H_{98}^* < 1.4$. The operational domain does not expand when ignoring the $H_{98}^* < 1.4$ requirement because of the assumed limit $C_\beta < 0.5$ (see Fig.~\ref{fig:scan1}.d), which implies the restriction $f_\mathrm{GW} \gtrsim 85\,\%$ along the narrow $P_\mathrm{fus} < 1.8$\,GW, $Q_\mathrm{sci} > 11$ domain. However, interpreting the estimated $C_\beta < 0.5$ as a soft limit (see discussions in Sec.~\ref{sec:mhd} for details), lower density regimes can be explored by reducing $f_\mathrm{GW}$ while increasing $\beta_\mathrm{N}$ along the narrow domain where the limits on fusion power and fusion $Q$ are both satisfied. As discussed in Sec.~\ref{sec:eccdnorm}, some scalings predict minimum confinement in lower density regimes ($f_\mathrm{GW} \lesssim 90$\,\%), meaning that there is a chance that these regimes are more accessible from a confinement point of view, supposing that any of these scalings are more valid than IPB98(y,2) in STEP relevant regimes. The two flat-top point categories of the EC only scenario are here labelled EC-HD~\cite{echd} and EC-LD~\cite{ecld} (for high density and low density, respectively). As discussed in Sec.~\ref{sec:scan2}, the EC+EBW scenario suggests a domain of relatively high EBW power ($P_\mathrm{EBW} \gtrsim 100$\,MW) when $H_{98}^* < 1.2$. Significantly lower $P_\mathrm{aux}$ and higher $Q_\mathrm{sci}$ domains can in principle be accessed if $\beta_\mathrm{N}$ is increased (to expand the domain with $P_\mathrm{fus} > 1.5$\,GW to lower $P_\mathrm{EBW}$) while relaxing the $H_\mathrm{98}^*$ limit. Two flat-top point categories for the EC+EBW scenario are here labelled as EB-CC~\cite{ebcc} (CC: conservative confinement) and EB-HQ~\cite{ebhq} (HQ: high fusion $Q$).

Higher fidelity JETTO modelling has explored these four operational points with more realistic physics assumptions compared to those of the scans in Section~\ref{sec:scans}. Rather than prescribing the radiation fraction, impurity radiation is self-consistently predicted with atomic physics from SANCO~\cite{sanco} combined with impurity transport from JETTO. The seeded radiating impurity species are xenon and argon, with argon primarily being used for divertor conditioning using gas puffing, whereas xenon is seeded with doped fuelling pellets to bring the core radiation to the required level for detachment access (see details in Sec.~\ref{sec:ploss}). The criteria for the optimisation of the $q$-profile shaping have also been refined when developing the candidate flat-top points compared to the template simulations (see~\cite{brown} for a full list of criteria). A summary of 0D outputs from the higher fidelity simulations are presented in Table~\ref{tab:cand}. With the radiation distribution self-consistently calculated from core impurities, this is taken into account when calculating the $H_{98}^*$ confinement factor (but not for $H_\mathrm{ITPA-IL}$). Each operating point has been further analysed with MHD code MARS for assessing RWM stability. Values for $\beta_\mathrm{N}^\mathrm{no{\mhyphen}wall}$ and $\beta_\mathrm{N}^\mathrm{ideal{\mhyphen}wall}$ have been computed with the code, rather than using the values that were used in the analysis of the scan results. The results are presented in Table~\ref{tab:mars}.

\begin{figure}\centering
\includegraphics[width=\textwidth]{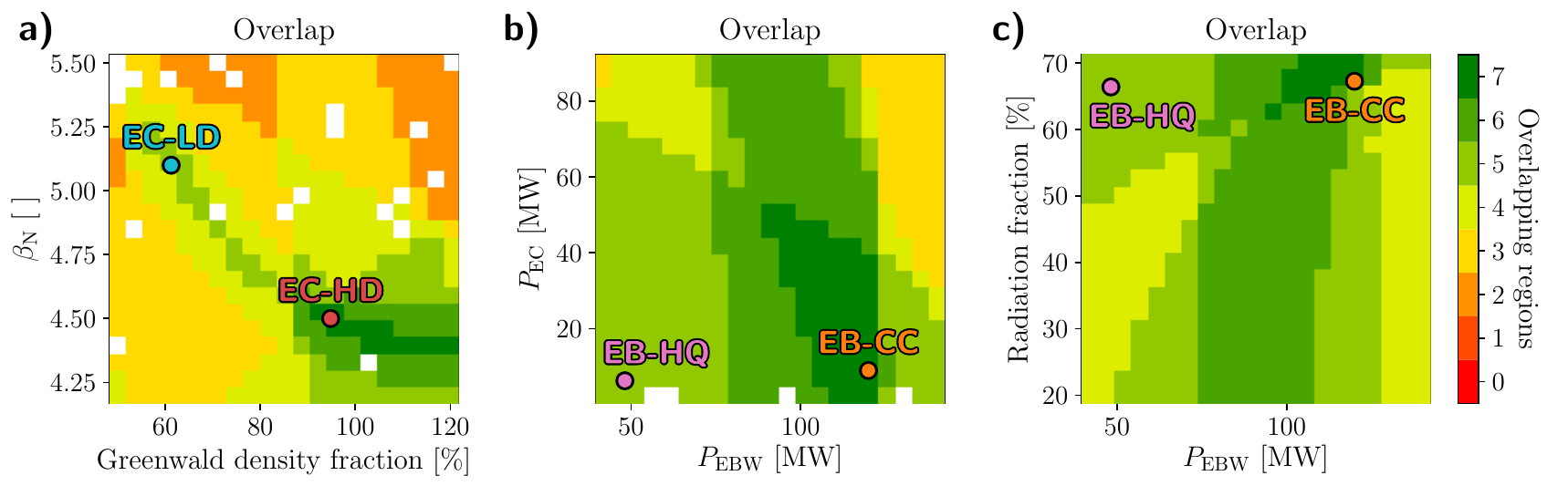}
\caption{The location of the four candidate flat-top operational points in the scans of Sections~\ref{sec:scan1}, \ref{sec:scan2} and \ref{sec:scan3}. The overlap plots of Figs.~\ref{fig:cand}.a, \ref{fig:cand}.b and \ref{fig:cand}.c correspond to Figs.~\ref{fig:scan1}.i, \ref{fig:scan2}.h and \ref{fig:scan3}.h, respectively. The two EC only operational points, EC-HD and EC-LD, are placed in the EC only scan (Fig.~\ref{fig:cand}.a), whereas the EC+EBW scenario points, EB-CC and EB-HQ, are in the two EC+EBW scans (Figs.~\ref{fig:cand}.b and \ref{fig:cand}.c).}\label{fig:cand}
\end{figure}

The operational points are placed on top of the overlap plots for the scans of Sections~\ref{sec:scan1}, \ref{sec:scan2} and \ref{sec:scan3} in Fig.~\ref{fig:cand}, to indicate where the points are in relation to the operational spaces of the scans. The impact of impurities on the plasma properties, as well as the self-consistent recalculation of the auxiliary heating and current drive efficiencies at the candidate operational points, mean that the points do not exactly match the corresponding locations in the scans. The EC auxiliary powers also differ for the EC-LD and the EB-HQ points relative to the scan templates used in the ($f_\mathrm{GW}$, $\beta_\mathrm{N}$) and the ($P_\mathrm{EBW}$, $f_\mathrm{rad}$) scans, respectively.

\begin{table}[ht]
    \centering
    \caption{Summary of 0D outputs from the candidate operational points. The top six rows show the outputs relative to defined operational limits indicated by colour, using the same colourmap as the 2D plots of Section~\ref{sec:scans} (see e.g. Figs.~\ref{fig:scan3}.a -- \ref{fig:scan3}.e and \ref{fig:scan3}.g). Complementary data on confinement, fuelling and burn-up is presented in Table~\ref{tab:conf}.\vspace{2mm}}
    \begin{tabular}{rlcccc}\hline
        & & \textbf{EC-HD} & \textbf{EC-LD} & \textbf{EB-CC} & \textbf{EB-HQ} \\\hline
        $P_\mathrm{fus}$ & [GW] & \textbf{\color{g0}1.68} &\textbf{\color{g0}1.51} &\textbf{\color{g0}1.56} &\textbf{\color{g0}1.62} \\
        $Q_\mathrm{sci}$ & [ ] & \textbf{\color{g0}11.1} &\textbf{\color{r0}9.22} &\textbf{\color{g0}12.0} &\textbf{\color{g0}29.1} \\
        $I_\mathrm{p}$ & [MA] & \textbf{\color{g0}21.2} &\textbf{\color{r0}25.1} &\textbf{\color{g0}22.7} &\textbf{\color{g0}21.3} \\
        $C_{\beta}$\footnotemark & [ ] & \textbf{\color{r0}0.54} &\textbf{\color{r0}0.60} &\textbf{\color{g0}0.11} &\textbf{\color{g0}0.27} \\
        $P_\mathrm{sep}/R_\mathrm{geo}$ & [MW/m] & \textbf{\color{g0}39.0} &\textbf{\color{g0}40.1} &\textbf{\color{g0}39.9} &\textbf{\color{g0}34.9} \\
        $H_{98}^*$ & [ ] & \textbf{\color{y0}1.35} &\textbf{\color{r0}1.62} &\textbf{\color{g0}1.19} &\textbf{\color{y0}1.36} \\
        $P_\mathrm{EC}$ & [MW] & 150 & 160 & 9.00 & 6.30 \\
        $P_\mathrm{EBW}$ & [MW] & --- & --- & 120 & 48.2 \\
        $P_\mathrm{rad}$ & [MW] & 343 & 321 & 297 & 252 \\
        $\langle n_\mathrm{e}\rangle$ & [10$^{19}$ m$^{-3}$] & 15.1 & 11.6 & 16.4 & 15.6 \\
        $\langle n_\mathrm{e}\rangle_\mathrm{line}/n_\mathrm{GW}$ & [\%] & 94.8 & 61.2 & 96.4 & 97.8 \\
        $\langle T_\mathrm{e}\rangle$ & [keV] & 11.0 & 18.1 & 9.69 & 10.1 \\
        $\langle T_\mathrm{i}\rangle/\langle T_\mathrm{e}\rangle$ & [ ] & 1.07 & 1.24 & 1.04 & 1.06 \\
        $I_\mathrm{BS}/I_\mathrm{p}$ & [\%] & 89.5 & 81.1 & 81.2 & 89.7 \\
        $I_\mathrm{EC}$ & [MA] & 1.87 & 4.22 & 0.155 & 0.126 \\
        $I_\mathrm{EBW}$ & [MA] & --- & --- & 3.95 & 1.70 \\
        $I_\mathrm{EC}/P_\mathrm{EC}$ & [kA/MW] & 12.5 & 26.4 & 17.2 & 20.0 \\
        $I_\mathrm{EBW}/P_\mathrm{EBW}$ & [kA/MW] & --- & --- & 32.9 & 35.2 \\
        $I_\mathrm{aux}/P_\mathrm{aux}$ & [kA/MW] & 12.5 & 26.4 & 31.8 & 33.4 \\
        $\beta_\mathrm{N}$ & [\%] & 4.50 & 5.10 & 3.93 & 4.30 \\
        $\beta_\mathrm{tor}$ & [\%] & 15.0 & 20.0 & 13.9 & 14.3 \\
        $B_\mathrm{axis}$ & [T] & 2.46 & 2.46 & 2.45 & 2.36 \\
        $l_\mathrm{i}$ & [ ] & 0.491 & 0.529 & 0.419 & 0.459 \\
        $l_\mathrm{i}(3)$ & [ ] & 0.275 & 0.296 & 0.234 & 0.257 \\
        $q_\mathrm{min}$ & [ ] & 2.50 & 2.15 & 2.56 & 2.42 \\
        $q_{95}$ & [ ] & 9.07 & 7.93 & 8.03 & 8.84 \\
        $Z_\mathrm{eff}$ & [ ] & 2.72 & 4.58 & 2.31 & 2.43 \\\hline
    \end{tabular}
    \label{tab:cand}
    \\\flushleft\footnotemark[\value{footnote}]{\footnotesize\hspace{.46em}$C_\beta$ is evaluated using values of $\beta_\mathrm{N}^\mathrm{no{\mhyphen}wall}$ and $\beta_\mathrm{N}^\mathrm{ideal{\mhyphen}wall}$ predicted by MARS~\cite{mars} for each individual operating point, see details of Table~\ref{tab:mars}.}
\end{table}
\begin{table}[ht]
    \centering
    \caption{0D output for the candidate operational points related to confinement, fuelling and burn-up. $\tau_\mathrm{E,100\%}^*$ is the energy confinement time with 100\,\% of the radiation subtracted, and $\tau_\mathrm{ei}$ is the energy exchange time. All of the listed confinement factors have been evaluated without radiation correction, for consistency.\vspace{2mm}}
    \label{tab:conf}
    \begin{tabular}{rlcccc}\hline
        & & \textbf{EC-HD} & \textbf{EC-LD} & \textbf{EB-CC} & \textbf{EB-HQ} \\\hline
        $H_\mathrm{IPB98(y,2)}$ & [ ] & 1.15 & 1.43 & 1.03 & 1.18 \\
        $H_\mathrm{ITPA20}$ & [ ] & 1.38 & 1.61 & 1.23 & 1.41 \\
        $H_\mathrm{ITPA20{\mhyphen}IL}$ & [ ] & 1.04 & 1.12 & 0.914 & 1.07 \\
        $H_\mathrm{Petty08}$ & [ ] & 0.646 & 0.817 & 0.599 & 0.701 \\
        $H_\mathrm{NSTX(19)}$ & [ ] & 0.306 & 0.364 & 0.299 & 0.343 \\
        $H_\mathrm{Kurskiev}$ & [ ] & 0.106 & 0.152 & 0.0965 & 0.111 \\
        $S_\mathrm{pel}$ & [10$^{20}$ s$^{-1}$] & 93.5 & 0.430 & 102 & 102 \\
        $f_\mathrm{BU}$ & [\%] & 6.39 & N/A & 5.41 & 5.68 \\
        $\tau_\mathrm{p}$ & [s] & 11.5 & N/A & 11.4 & 11.0 \\
        $\tau_\mathrm{E}$ & [s] & 1.24 & 1.67 & 1.30 & 1.53 \\
        $\tau_\mathrm{E,100\%}^*$ & [s] & 4.20 & 5.41 & 3.98 & 4.55 \\
        $\tau_\mathrm{ei}$ & [s] & 1.99 & 4.91 & 1.54 & 1.56 \\\hline
    \end{tabular}
\end{table}
\begin{table}[ht]
    \centering
    \caption{Evaluation of the RWM stability limits with MARS~\cite{mars} MHD code for each of the four candidate flat-top points. The destabilisation parameter $C_\beta$ is defined in eq.~(\ref{eq:cbdef}).\vspace{2mm}}
    \label{tab:mars}
    \begin{tabular}{lcccc}\hline
         & \textbf{EC-HD} & \textbf{EC-LD} & \textbf{EB-CC} & \textbf{EB-HQ} \\\hline
         $\beta_\mathrm{N}$ & 4.50 & 5.10 & 3.93 & 4.30 \\
         $\beta_\mathrm{N}^\mathrm{no{\mhyphen}wall}$ & 3.31 & 3.92 & 3.72 & 3.75 \\
         $\beta_\mathrm{N}^\mathrm{ideal{\mhyphen}wall}$ & 5.53 & 5.90 & 5.72 & 5.76 \\
         $C_\beta$ & 0.54 & 0.60 & 0.11 & 0.27 \\\hline
    \end{tabular}
\end{table}
\begin{figure}
    \centering
    \includegraphics[width=\textwidth]{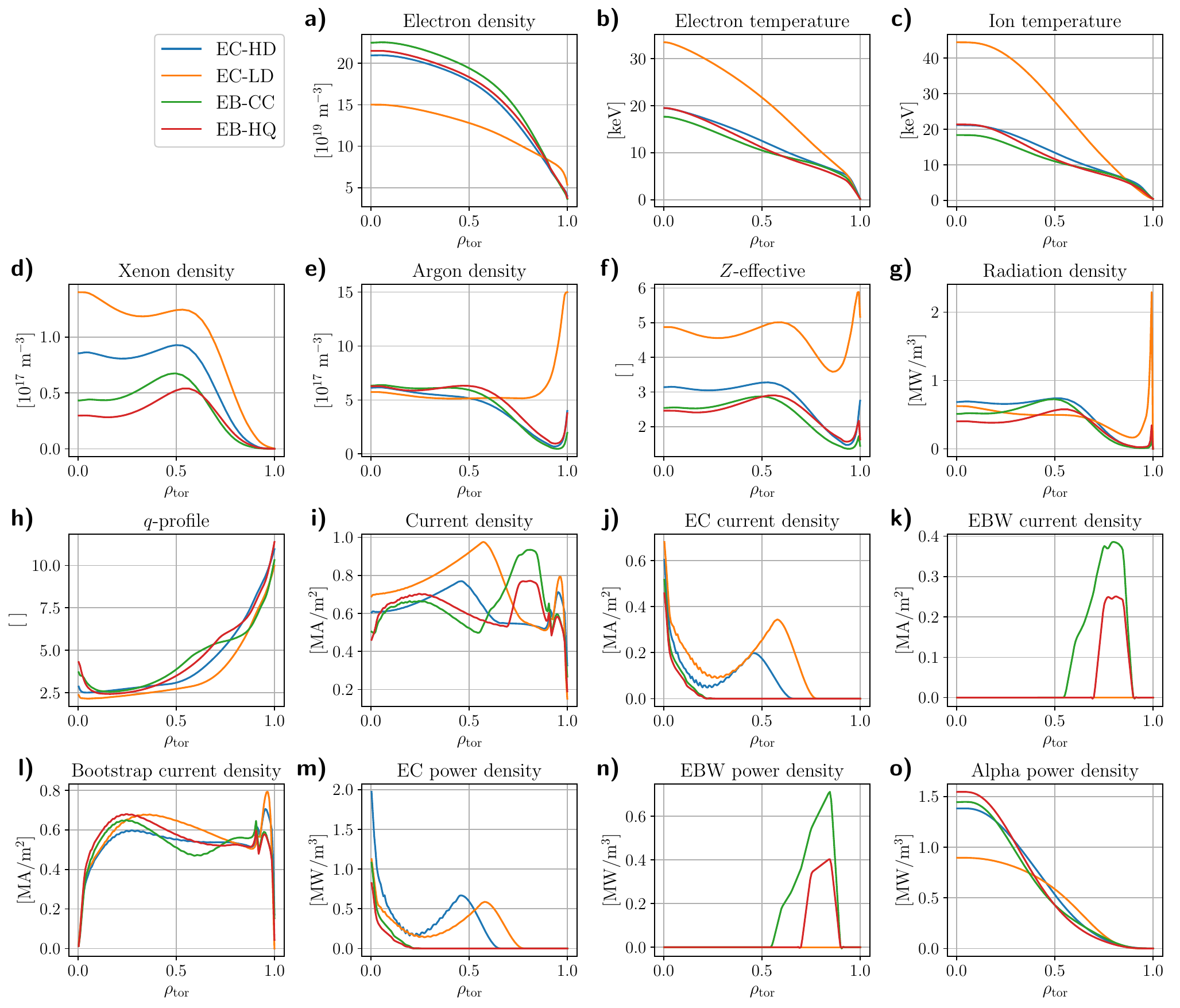}
    \caption{Profile data for candidate operational points.}
    \label{fig:cprof}
\end{figure}

EC-HD satisfies all the listed operational limits except for $C_\beta < 0.5$, assuming that the optimistic confinement assumption is valid. The $C_\beta$ limit is only marginally exceeded at $C_\beta = 0.54$, and it is possible that RWM control is still manageable with a combination of induced rotation and external control coils, as discussed in Section~\ref{sec:mhd}. As indicated by Fig.~\ref{fig:s11d}.d, $C_\beta$ can be reduced either by reducing $\beta_\mathrm{N}$ or by increasing $f_\mathrm{GW}$. It is likely that a reduction of $\beta_\mathrm{N}$ will also reduce $Q_\mathrm{sci}$ below 11. An increased Greenwald fraction might also reduce the required degree of confinement, as indicated by Figs.~\ref{fig:hfactor}.a -- \ref{fig:hfactor}.f ($H_\mathrm{NSTX(19)}$ is the only exception). As discussed in Section~\ref{sec:mhd}, operation at densities above the Greenwald limit has been demonstrated in scenarios with low ELM activity~\cite{ngw}. Whether these regimes are applicable to STEP without significant risk of disruption events remains to be analysed.

The low density EC operational point is out of range with respect to the suggested operational limit in $Q_\mathrm{sci}$, with $Q_\mathrm{sci} = 9.22 < 11$. In addition, the EC-LD point has a plasma current $I_\mathrm{p} = 25.11\,\textrm{MA} > I_\mathrm{p,max} = 25\,\textrm{MA}$ (the limit of 23\,MA was assumed for the scans only, see discussion in Section~\ref{sec:cur}), and $C_\beta = 0.60 > C_{\beta,\mathrm{max}} = 0.5$. The scan of Sec.~\ref{sec:scan1} (scan parameters $f_\mathrm{GW}$ and $\beta_\mathrm{N}$ in EC only scenario) indicates that the most efficient ways to increase $Q_\mathrm{sci}$ is to either increase $\beta_\mathrm{N}$ or $f_\mathrm{GW}$. Increasing $\beta_\mathrm{N}$ only is likely not to access an operational point, as there is a positive trend of $I_\mathrm{p}$ with respect to $\beta_\mathrm{N}$, and $I_\mathrm{p}$ of EC-LD already exceeds the suggested limit of 25\,MA. Increasing the Greenwald density fraction could potentially decrease $I_\mathrm{p}$ due to the non-monotonic dependency of $I_\mathrm{p}$ with respect to $f_\mathrm{GW}$ (see Fig.~\ref{fig:s11d}.c). However, the minimum of $I_\mathrm{p}$ in that dependency is expected to be close to the EC-LD point ($f_\mathrm{GW} = 61.23$\,\%, $\beta_\mathrm{N} = 5.10$), meaning that a significant decrease of $I_\mathrm{p}$ is not expected. A combination of increased $f_\mathrm{GW}$ and decreased $\beta_\mathrm{N}$ is likely required to access an operational point relative to EC-LD in terms of $Q_\mathrm{sci}$ and $I_\mathrm{p}$. Moving in this direction in parameter space is also likely to reduce $C_\beta$, as indicated by Fig.~\ref{fig:s11d}.d. An issue with the high on-axis current density from ECCD in the EC-LD point is that $q_\mathrm{min} = 2.15$ is below the recommended lower limit of $2.2$. Simply reducing the EC power is not viable since it would decrease the fusion power, which is already close to the lower limit of 1.5\,GW.

One of the main motivations for the low-density EC operating point was to minimise the confinement according to empirical scalings different from IPB98(y,2). However, as seen in Table~\ref{tab:conf}, all of the listed confinement factors predict the highest value for the EC-LD point, including the $H_\mathrm{ITPA20{\mhyphen}IL}$ scaling. It is possible that $H_\mathrm{ITPA20{\mhyphen}IL}$ can be reduced by increasing the plasma current, due to its relatively strong dependence in $\tau_\mathrm{ITPA20{\mhyphen}IL} \propto I_\mathrm{p}^{1.291}$. However, this is not an option due to the current already exceeding tolerable levels, as outlined in the paragraph above. In addition to the confinement factors, Table~\ref{tab:conf} also presents data on the pellet fuelling rate $S_\mathrm{pel}$, the fusion burn-up fraction $f_\mathrm{BU}$, and the particle confinement time $\tau_\mathrm{p} = V\langle n_\mathrm{e}\rangle/S_\mathrm{pel}$. With the relatively low pellet fuelling rate of the EC-LD point, the burn-up fraction and the particle confinement time become unphysically large (of the orders 1\,000\,\% and 1\,000\,s, respectively). It is still to be confirmed whether this is a result of the confinement assumption in the modelling, the way that the confinement time and burn-up fraction are evaluated, or a combination of both. It is evident that more work is needed to develop a potentially viable alternative for the EC scenario in the low-density regime.

The most promising operational point of the four candidates is EB-CC, which operates within all of the suggested limits, including the conservative confinement assumption $H_{98}^* < 1.2$. As discussed in Section~\ref{sec:ecebw}, the main caveat of the EC+EBW scenario is the assumption of the high current drive efficiency of EBW, which has little experimental verification. The EB-HQ point is significantly closer to reactor relevant conditions, with $Q_\mathrm{sci} = 29.07$. The relatively low $P_\mathrm{sep}/R_\mathrm{geo} = 34.87$\,MW/m also means that detachment access is expected to be easier compared to the other candidate flat-top points. Even though $\beta_\mathrm{N}$ had to be increased to 4.3 to reach target fusion powers, which is above the estimated no-wall $\beta_\mathrm{N}$ limit of 3.75 (see Table~\ref{tab:mars}), $C_\beta = 0.27$ is still below the suggested 0.5 limit.

Observing the profile data for the four operational points in Fig.~\ref{fig:cprof}, all outputs are relatively similar, except for the EC-LD point, which is characterised by low density, high temperature, high $T_\mathrm{i}/T_\mathrm{e}$, high $Z_\mathrm{eff}$, and high argon edge density and radiation in comparison. As seen in Table~\ref{tab:conf}, the energy exchange time $\tau_\mathrm{ei}$ is similar to the energy confinement time $\tau_\mathrm{E}$ without radiation correction for all points except EC-LD, which has $\tau_\mathrm{ei}$ closer to $\tau_\mathrm{E,100\%}^*$. The relatively longer energy exchange time of EC-LD allows for electron and ion temperatures farther from equilibration. The fact that $T_\mathrm{i} > T_\mathrm{e}$ rather than the other way around is due to a combination of $\chi_\mathrm{i} \ll \chi_\mathrm{e}$ and the overall higher temperatures, which causes a larger ion heating fraction of the alpha heating. The higher edge argon density of EC-LD is a consequence of the difference in neoclassical transport that follows from the different pedestal. The current densities of the two EC points, EC-HD and EC-LD, are flat for $\rho_\mathrm{tor} \lesssim 0.4$ and $\rho_\mathrm{tor} \lesssim 0.6$, respectively, compared to the two EC+EBW operational points. The flatness of the current density profile is due to optimisation of the EC current density profile, which is adapted to be complementary to the bootstrap current in such a way that the current density is nearly constant in the domain where EC waves couple to the plasma (compare Figs.~\ref{fig:cprof}.j and \ref{fig:cprof}.l). The resulting $q$-profile is also flat, sitting at a value far from low-order rationals across most of the minor radius, which is beneficial for avoiding destabilisation of MHD modes such as neoclassical tearing modes.

\section{Conclusions}\label{sec:conc}

The flat-top operational space of the STEP (Spherical Tokamak for Energy Production) experiment has been explored using JETTO integrated core plasma model in the initial stages of the development of a net energy output scenario. The modelling focuses on stationary non-inductive flat-top operation points rather than whole pulses. The operational space is explored by performing a series of scans in key plasma parameters, where each scan point is a JETTO simulation that runs until 0D output parameters reach convergence. The output parameters are studied in relation to a set of operational limits that provide guidelines to the viable scenarios. The operational limits relate to e.g. fusion gain, plasma stability and control, detachment access and exhaust management, engineering limitations, and viability of the modelling assumptions. By studying the operational spaces indicated by the scans, a set of candidate flat-top operational points have been developed.

Two main types of scenarios are studied in terms of auxiliary heating and current drive systems. The first type runs with EC (electron cyclotron) systems only, whereas the second type of scenarios uses a combination of EC and EBW (electron Bernstein waves) systems. Within the required operational domains, the EC systems only provide absorption in the central part of the plasma ($\rho_\mathrm{tor} \lesssim 0.5$), whereas the EBW systems can access off-axis heating and current drive. Even though the auxiliary systems and $\alpha$-particle heating primarily heat the electrons, ion temperatures comparable to electron temperatures are still expected due to the assumption that $\chi_\mathrm{e} \gg \chi_\mathrm{i}$\footnote{We note that first nonlinear gyrokinetic simulations for STEP also find that the heat fluxes $Q_\mathrm{e} \gg Q_\mathrm{i}$~\cite{giacomin}.} (based on experience in spherical tokamak plasmas~\cite{kaye}). Extrapolation to large size tokamaks also predict improved equilibration of electron and ion temperatures in primarily electron heated plasmas~\cite{angioni}.

In order to access a wide range of densities and temperatures in the scans, the Greenwald density fraction, $f_\mathrm{GW}$, and the normalised beta, $\beta_\mathrm{N}$, were set-up as inputs to the modelling. Feedback systems acting on the fuelling rate of the plasma and on rescaling of the anomalous diffusivities were applied to access the set targets in $f_\mathrm{GW}$ and $\beta_\mathrm{N}$, respectively. The confinement enhancement factor relative to the IPB98(y,2) scaling, $H_\mathrm{IPB98(y,2)}$, provides an indicator as to the viability of achieving the required confinement, with $H_\mathrm{IPB98(y,2)} \gg 1$ indicating an overestimated energy confinement. In mapping out the available operating space, different degrees of confinement have been assumed to account for uncertainties in the expected transport for STEP relevant regimes. A lower confinement assumption tend to require high-density, high auxiliary power operating points, whereas better confinement would allow access to broader density and power parameter regimes.

Four flat-top points were suggested, two points that were using EC heating systems, and two points that used a combination of EC and EBW heating. The two EC operating points primarily differ by density, with the high-density point having a more restrictive confinement assumption compared to the low-density point. The low-density point has to operate at significantly higher temperatures than the high-density point in order to reach a fusion power of 1.5\,GW (the lower recommended limit for STEP). However, it only reaches a scientific fusion energy gain factor $Q_\mathrm{sci}$ of 9.2, whereas the high-density point predicts operation at $Q_\mathrm{sci} = 11.1$. Other performance issues of the low-density point that have been demonstrated is a higher risk of uncontrollable growth of resistive wall modes and a high plasma current $I_\mathrm{p} > 25$\,MA, which might cause unacceptable levels of damage to the tokamak during disruption events.

The two EC+EBW scenario points mainly differ by the auxiliary power, with the high power case using $P_\mathrm{EC} = 9.0$\,MW and $P_\mathrm{EBW} = 120.0$\,MW, compared to $P_\mathrm{EC} = 6.3$\,MW and $P_\mathrm{EBW} = 48.2$\,MW for the low power case. The high-power EC+EBW scenario point has a much lower confinement factor than both of the EC points, whereas the low-power point has a similar confinement factor as the one estimated for the high-density EC point, making the EC+EBW scenario in general more viable than the EC scenario in terms of confinement assumptions. The low-power EC+EBW point also has relatively high $Q_\mathrm{sci} = 29.1$, making it an attractive point for the development of commercial fusion reactors. 

The main caveat of the EC+EBW scenario in general is that available EBW heating and current drive models have limited experimental validation, making the assumption of a high EBW current drive efficiency ($I_\mathrm{EBW}/P_\mathrm{EBW} > 30$\,kA/MW) less certain. On the other hand, a scenario with only EC heating and current drive requires more optimistic confinement time assumptions in order to provide sufficient fusion gain. Since transport models also lack experimental validation for STEP relevant regimes, considerable uncertainty remains in the viability of both scenarios. However, it should be noted that one of the most important products of the presented work is not necessarily the specific operating points that have been derived, but rather the development of a robust methodology for generating designs of non-inductive, burning plasmas. On a related note, the spherical tokamak MAST-U will soon undergo significant enhancements, including the installation of an EBW system and a second neutral beam injector box. With these upgrades, MAST-U will be able to provide important experimental data that will improve our understanding of EBW heating and current drive and of transport in more STEP relevant plasmas.

\section*{Acknowledgements}
This work has been funded by STEP, a UKAEA programme to design and build a prototype fusion energy plant and a path to commercial fusion. To obtain further information on the data and models underlying this paper please contact PublicationsManager@ukaea.uk.

\section*{References}
\bibliographystyle{iopart-num}
\providecommand{\newblock}{}

\end{document}